\newcommand{\bs}{{\bf \widehat{s}}}
\newcommand{\bb}{\boldsymbol{b}}
\newcommand{\nVis}{N_{\text{vis}}}
\newcommand{\kperp}{k_\perp}
\newcommand{\by}{{\bf y}}
\newcommand{\ftN}{\mathsf{\widetilde{N}}}
\newcommand{\bftN}{\boldsymbol{\ftN}}
\newcommand{\fc}{\widetilde{c}}
\newcommand{\Lm}{\mathsf{\Lambda}}
\newcommand{\bLm}{\boldsymbol{\Lm}}
\newcommand{\Pm}{\mathsf{\Psi}}
\newcommand{\Pmt}{\Pm^\intercal}
\newcommand{\Lmt}{\Lm^\intercal}
\newcommand{\bPm}{\boldsymbol{\Pm}}
\newcommand{\bPmt}{\bPm^\intercal}
\newcommand{\bLmt}{\bLm^\intercal}
\newcommand{\bc}{{\bf c}}
\newcommand{\ftC}{\mathsf{\widetilde{C}}}
\newcommand{\bftC}{\boldsymbol{\ftC}}

\newcommand{\bS}{\boldsymbol{\mathsf{S}}}

\newcommand{\ftV}{\widetilde{V}}
\newcommand{\measPhi}{M^\phi}
\newcommand{\measEta}{M^\eta}
\newcommand{\Neff}{n_{\text{eff}}}
\newcommand{\bW}{\boldsymbol{\mathsf{W}}}

\newcommand{\bA}{\boldsymbol{\mathsf{A}}} 

\newcommand{\sMin}{S_{\text{min}}}
\newcommand{\sMax}{S_{\text{max}}}

\newcommand{\freq}{\nu} 

\newcommand{\bAt}{\boldsymbol{\mathsf{A}}^\intercal}

\newcommand{\cN}{\mathsf{N}}
\newcommand{\cC}{\mathsf{C}}
\newcommand{\bN}{ \boldsymbol{\cN}}
\newcommand{\bu}{{\bf  u}}
\newcommand{\bk}{{\bf k}}
\newcommand{\br}{{\bf r}}
\newcommand{\ftR}{\mathsf{\widetilde{R}}}
\newcommand{\bftR}{\boldsymbol{\ftR}}
\newcommand{\ftS}{\mathsf{\widetilde{S}}}
\newcommand{\bftS}{\boldsymbol{\ftS}}

\newcommand{\feta}{\widetilde{\eta}}
\newcommand{\fphi}{\widetilde{\phi}}

\newcommand{\re}{\text{Re}}

\newcommand{\im}{\text{Im}}
\newcommand{\hg}{\widehat{g}}
\newcommand{\cov}{\text{Cov}}
\newcommand{\bn}{\boldsymbol{n}}
\newcommand{\covC}{\mathsf{C}}
\newcommand{\bC}{\boldsymbol{\covC}}
\newcommand{\bB}{\boldsymbol{\mathsf{B}}}

\newcommand{\bDs}{\boldsymbol{\Delta} \boldsymbol{\widehat{s}}}
\newcommand{\bPhi}{\boldsymbol{\phi}}
\newcommand{\bEta}{\boldsymbol{\eta}}

\newcommand{\bmeasEta}{{\bf M^\eta}}
\newcommand{\bmeasPhi}{{\bf \measPhi}}
\newcommand{\bdwEta}{\boldsymbol{\widehat{\eta}}}
\newcommand{\bdwPhi}{\boldsymbol{\widehat{\phi}}}
\newcommand{\bBt}{\boldsymbol{\mathsf{B}}^\intercal}
\newcommand{\trans}{^\intercal}
\newcommand{\cEta}{\mathsf{C_\eta}}
\newcommand{\bcEta}{\boldsymbol{\cEta}}

\newcommand{\cPhi}{\mathsf{C_\phi}}
\newcommand{\bcPhi}{\boldsymbol{\cPhi}}

\newcommand{\cR}{\mathsf{R}}

\newcommand{\nAnt}{N_{\text{ant}}}
\newcommand{\kparmin}{k_\parallel^{\text{min}}}
\newcommand{\sigmaAnt}{\sigma_{\text{ant}}}
\newcommand{\kpar}{k_\parallel}

\newcommand{\dAnt}{d_{\text{ant}}}

\newcommand{\bR}{\boldsymbol{\mathsf{R}}}
\newcommand{\sigmaBeam}{\sigma_b}

\newcommand{\varR}{\sigma_r^2(\sMin)}

\documentclass[useAMS,usenatbib]{mnras}
\usepackage{amsmath,amssymb,aas_macros,graphicx,comment}
\begin{document}

\author[Ewall-Wice et al.]{Aaron Ewall-Wice$^{1,2}$ \thanks{E-mail: aaronew@mit.edu},
Joshua S. Dillon$^{3,4}$,
Adrian Liu$^{3,\dagger}$,
Jacqueline Hewitt$^{1,2}$,
\\
$^{1}$MIT Kavli Institute for Astrophysics and Space Research, Cambridge, MA 02139, USA \\
$^{2}$Dept. of Physics and MIT Kavli Institute, Massachusetts Institute of Technology, Cambridge, MA 02139, USA \\
$^{3}$Dept. of Astronomy and Radio Astronomy Laboratory, UC Berkeley, Berkeley CA 94720\\
$^{4}$Berkeley Center for Cosmological Physics, University of California, Berkeley, Berkeley, CA 94720\\
$^{\dagger}$Hubble Fellow. \\
}

\title[Calibration Modeling Errors in 21\,cm Power Spectra]{The Impact of Modeling Errors on Interferometer Calibration for 21\,cm Power Spectra}
\maketitle
\begin{abstract}
We study the impact of sky-based calibration errors from source mismodeling on 21\,cm power spectrum measurements with an interferometer and propose a method for suppressing their effects. While emission from faint sources that are not accounted for in calibration catalogs is believed to be spectrally smooth, deviations of true visibilities from model visibilities are not, due to the inherent chromaticity of the interferometer's sky-response (the ``wedge"). Thus, unmodeled foregrounds, below the confusion limit of many instruments, introduce frequency structure into gain solutions on the same line-of-sight scales on which we hope to observe the cosmological signal. 
We derive analytic expressions describing these errors using linearized approximations of the calibration equations and estimate the impact of this bias on measurements of the 21\,cm power spectrum during the Epoch of Reionization (EoR). Given our current precision in primary beam and foreground modeling, this noise will significantly impact the sensitivity of existing experiments that rely on sky-based calibration.  Our formalism describes the scaling of calibration with array and sky-model parameters and can be used to guide future instrument design and calibration strategy. We find that sky-based calibration that down-weights long baselines can eliminate contamination in most of the region outside of the wedge with only a modest increase in instrumental noise.  
\end{abstract}
\section{Introduction}
Observations of redshifted 21\,cm emission are poised to unveil the properties of the earliest luminous sources in the universe, their impact on the global state of the intergalactic medium and how they affected the subsequent generations of stars and galaxies (see \citealt{McQuinn:2015,Furlanetto:2016Review} for recent reviews). 
	
One approach to detecting the cosmological 21\,cm signal is to measure the fluctuations in the brightness temperature which can be mapped tomographically with a radio interferometer.  To enhance the significance of a detection, most experiments are attempting to measure the spherically averaged power spectrum of these fluctuations. The mitigation of foregrounds that are four to five orders of magnitude brighter than the signal itself is a central challenge that 21\,cm experiments must overcome but is greatly aided by the spectral smoothness of these foregrounds \citep{DiMatteo:2002,Oh:2003,Morales:2004,Zaldarriaga:2004}. An interferometer measures the brightness distribution on the sky by cross correlating the outputs from many pairs of antennas. Flat-spectrum radio waves from a single point source, at a given time of observation, appear at a fixed time delay in the correlation between two antennas. Since the delay between two correlated antenna outputs is the Fourier dual to frequency, each fixed-delay source introduces a sinusoidal ripple as a function of frequency with a period that is inversely proportional to the difference of the arrival times of that source at the two correlated antennas. This sinusoid in frequency will correspond to a single comoving cosmological mode. In the absence of reflections, the maximal delay between signals arriving from a source on the sky (corresponding to the maximal line-of-sight (LoS) cosmological Fourier mode that is contaminated) occurs when the source is located along the separation of the antennas, at the horizon. Hence, as viewed by an interferometer, the spectrally smooth foregrounds are naturally contained within a region of Fourier space known as the {\it wedge} \citep{Datta:2010,Vedantham:2012,Parsons:2012a,Morales:2012,Thyagarajan:2013,Liu:2014a,Liu:2014b}
which is given by the horizon delay for each baseline separation and increases with that separation.
 
It is also possible for the signal chain of the instrument to imprint spectral structure into the measured visibilities. For example, a reflection within the signal path can delay the correlated signal.  Hence, longer delays in the signal path contaminate finer frequency scales and are capable of leaking significant power outside of the wedge \citep{EwallWice:2015a,EwallWice:2016,Beardsley:2016b}.  Digital artifacts can also introduce fine spectral features such as those introduced by the polyphase filter bank on the Murchison Widefield Array (MWA) \citep{Offringa:2016}. For each delay that is contaminated by structure in the antenna gains, an attenuated copy of the foregrounds, which are $\sim 10^4-10^5$ times larger than the signal, is introduced. Using foreground simulations, \citet{Thyagarajan:2016} establish that in order to avoid contaminating the comoving LoS scales of several $h^{-1}$Mpc or smaller, which are targeted by 21\,cm experiments,    instrumental chromaticity beyond a $250$\,ns delay must be suppressed to the $\approx -50$\,dB level. Thus (a), the smoothness of the instrumental gain must meet this specification, or (b), calibration methods must be capable of suppressing any instrumental spectral structure to be within these limits.

Interferometric experiments have taken several distinct approaches to calibrating out instrumental spectral structure. Experiments focusing on imaging, such as the Murchison Widefield Array (MWA) \citep{Tingay:2013a}, the Low Frequency Array (LOFAR) \citep{VanHaarlem:2013}, early deployments of the Precision Array for Probing the Epoch of Reionization (PAPER) \citep{Jacobs:2011,Jacobs:2013,Kohn:2016} and the Giant Metrewave telescope (GMRT) \citep{Pagica:2013}\footnote{In the GMRT's case, a single, well known, pulsar is used while the rest of the sky is eliminated by difference time-steps that correspond to the pulsar's ``on" and ``off" states.} calibrate their gains on a model of the sky that is usually iteratively improved with self-calibration (where observed sources are fed into an updated sky-model which is used to obtain more accurate gain solutions). Pipelines such as the MWA's real time system (RTS), \citep{Mitchell:2008}, Fast Holographic Deconvolution (FHD) \citep{Sullivan:2012}, and {\tt sageCAL} \break \citep{Kazemi:2011,Kazemi:2013a,Kazemi:2013b}, rely on the modeling approach which we refer to as sky-based calibration. An alternative route is to constrain the instrumental gains using many redundant measurements of the same visibility \citep{Wieringa:1992,Liu:2010,Zheng:2014}. This strategy was implemented by the MIT EoR (MITEoR) array \citep{Zheng:2014,Zheng:2016a,Zheng:2016b}, the latest configurations of PAPER \citep{Parsons:2014,Ali:2015}, and the Hydrogen Epoch of Reionization Array (HERA) \citep{DeBoer:2016,Dillon:2016} that is now being commissioned in South Africa\footnote{HERA is designed to be fully
redundantly calibratable but it is
useful to assess the performance of sky-based calibration as an alternative with potentially different systematics. Since redundant calibration does not rely as much on a model of the sky, there exists the possibility of this array outperforming any of the predictions in this paper.}. Finally, {\it in-situ} calibration can be obtained using the injection of known signals \citep{Patra:2015}. The Canadian Hydrogen Intensity Mapping Experiment (CHIME) \citep{Newburgh:2014} is employing a  combination of redundant calibration, signal injection, and pulsar holography to correct for instrumental gains. 

Recent analyses of MWA data, using sky-based calibration have been contaminated by intrinsic chromaticity in the signal chain at the $\lesssim -20$\,dB level \citep{Dillon:2015b,EwallWice:2015a,Jacobs:2016,Beardsley:2016b} out to less than a delay of  $2\times10^3$\,ns, arising from a combination of  reflections in the beam-former to receiver cables and digital artifacts. Increasing the frequency degrees of freedom within sky-based calibration is a potential solution as the gains are permitted to absorb fine-scale instrumental frequency structure at high delays \citep{Offringa:2016} and improvements in features such as cable reflections were noted in power spectra calibrated with additional parameters \citep{Trott:2016}.

 While calibration solutions with fine frequency degrees of freedom are able to model the detrimental spectral features in an instrumental bandpass, they are susceptible to absorbing the imperfections in any sky-model used for calibration. Naively, errors in a smooth foreground model should not impart spectrally complex errors into a gain solution. However, because every gain participates in many baselines with varying lengths and (due to the wedge) intrinsic chromaticities, calibration can imprint the frequency-dependent errors of the longest baselines in which an antenna participates into its gain solution. The application of this gain solution on the short baselines that the antenna participates in will mix contamination from long to short baselines, potentially contaminating the EoR window. Recent studies by \citet{Barry:2016} (henceforth B16) and \citet{Patil:2016} have demonstrated the existence of these errors in simulations of the special cases of the MWA and LOFAR with specific point source realizations. It has not yet been established how these errors scale with the properties of the instrument and the source catalog and whether they will pose a fundamental limitation to upcoming 21\,cm experiments that expect to rely on sky-based calibration such as the Square Kilometre Array (SKA). Although B16 proposes a low-order-polynomial-based method to mitigate these effects, it generally relies upon intrinsically spectrally-smooth antenna bandpasses, which may not be the case for many interferometers.

In this paper, we employ linearized approximations of the calibration equations developed by \citet{Wieringa:1992} (W92) and \citet{Liu:2010} (L10) to investigate the amplitude of errors arising from incomplete calibration catalogs. Since these faint unmodeled sources can be described statistically \citep{Liu:2011,Liu:2011b,Trott:2012,Dillon:2013,Dillon:2014,Dillon:2015a,Trott:2016}, we will address the ensuing errors as a type of correlated noise which we will hereafter refer to as {\it modeling noise}. Unlike its thermal counterpart, modeling noise does not integrate down with observing time, biasing any power spectrum estimate. Since interleaved times in this noise are correlated, this bias cannot be eliminated (unlike thermal noise) by the technique of cross-multiplying interleaved time integrations (e.g.  \citealt{Dillon:2014}). We will derive equations describing the amplitude of modeling noise and its dependence on the properties of a radio interferometer such as the antenna count, distribution, and element size along with the depth of the calibration catalog. We use these equations to approximate the level of modeling noise in the existing MWA and LOFAR experiments (finding that our analytic results are in broad agreement with the simulation results in B16) along with the expected contamination in the upcoming instruments SKA-1 LOW and HERA. This contamination arises fundamentally from the chromaticity on long baselines, hence it can be eliminated by down-weighting long baselines in calibration, a strategy that we develop and verify in this paper. 

We take an analytic approach in order to illuminate the origins of modeling noise in 21\,cm power spectrum measurements and guide future array design and calibration strategies. For analytic tractability, we make a number of assumptions, which we attempt to describe clearly in the text, but do not necessarily hold for all observing scenarios. Thus, our quantitative results should be understood as accurate only to within an order of magnitude, illustrating how modeling noise scales with the properties of the sky catalog and instrumental parameters. Relaxing the assumptions in this paper for more accurate predictions is the subject of ongoing simulation work. 

This paper is organized as follows. In \S~\ref{sec:Formalism}, we introduce our analytic framework, based on W92 and L10, for describing the impact of calibration modeling errors on the 21\,cm power spectrum and discuss its dependence on array and catalog properties. In \S~\ref{sec:Results} we apply this formalism to predict the amplitude of calibration errors relative to 21\,cm fluctuations in current and upcoming experiments given our current knowledge of foregrounds and primary beams. In \S~\ref{sec:Mitigation}, we explore a strategy for eliminating this noise through inverse baseline-length weighting. We conclude in \S~\ref{sec:Conclusion}.
 
\section{Formalism}\label{sec:Formalism}
In this paper, we model gain errors as a statistical noise arising from the myriad of faint unmodeled sources. Such sources are not precisely modeled in calibration.

Since baselines and visibilities are formed from two antennas, we index them with a greek index and we index antennas with lower-case latin indices. We will also sometimes explicitly write a baseline index as a 2-tuple of antenna indices (e.g. $\alpha = (i,j)$). We describe the residual, $c_\alpha(\freq)$, between the true visibility formed from antennas $i$ and $j$, $v_\alpha^{\text{true}}(\freq)$, and the model visibility, $y_\alpha(\freq)$ as a random variable with a mean $\langle c_\alpha(\freq) \rangle$ and covariance $\cC_{\alpha \beta}(\freq,\freq')\equiv \left\langle \left[c_\alpha(\freq) - \langle c_\alpha(\freq) \rangle \right] \left[c_\beta(\freq') - \langle c_\beta(\freq') \rangle \right]^* \right \rangle$.  We assume that $c_\alpha$ is composed of the sum of  the 21\,cm signal, $s_\alpha$, unmodeled foregrounds, $r_\alpha$, and a component arising from thermal noise, $n_\alpha$ whose impact on calibration is explored in \citet{Trott:2016b}. The true visibility is the sum between the modeled and unmodeled component. 
\begin{equation}\label{eq:Components}
v_\alpha^{\text{true}}= y_\alpha + c_\alpha = y_\alpha + r_\alpha  + n_\alpha+ s_\alpha
\end{equation}
Each unmodeled component is statistically independent which means that
\begin{equation}
\cov[\bc,\bc^\dagger] \equiv \bC = \bR + \bN + \bS,
\end{equation}
where ${}^\dagger$ denotes the conjugate transpose of a vector or matrix. 
 In \S~\ref{ssec:Sources} we discuss expressions for the amplitude and frequency coherence of $r_\alpha$ in terms of a parameterized point source population and diffuse galactic emission. This ``noise" will be imprinted on the calibration solutions in a way that, for sufficiently small errors, is analytically tractable and can be described using the matrix formalism of W92 and L10, which we overview in \S~\ref{ssec:Linearized}. We derive expressions for the impact of these errors on the 21\,cm power spectrum in \S~\ref{ssec:PowerSpectrum} and the degree to which each visibility covariance contributes in \ref{ssec:Individual}. Using the expressions we derive, we discuss the scaling of modeling noise with the properties of the source-catalog and array in \S~\ref{ssec:Scaling}. Since both the 21\,cm signal and thermal noise terms are already well considered in the literature (W92, L10, \citealt{Trott:2016b}), we will focus on the contribution from $r_\alpha$. 

\subsection{The Statistics of Unmodeled Source Visibilities}\label{ssec:Sources}
Extensive work exists on statistical models of faint point sources in the power spectrum (e.g. \citealt{Wang:2006,Liu:2011,Trott:2012,Dillon:2013,Dillon:2015b,Trott:2016}) and we take an approach similar to these papers and assume the sources have uniform spectral structure (described by a single power law) that can be factored out of the visibilities and is far less significant than the frequency dependence introduced by the interferometric point-spread function. We now give an overview of our characterization of the unmodeled point sources along with the diffuse emission from the Galaxy. Since residual Galactic emission is, for the most part, uncorrelated with residual point-source emission, the covariance of unmodeled emission on each baseline is given by the sum of the covariance of each source,
\begin{equation}
\bR = \bR^P + \bR^G,
\end{equation}
where $\bR^P$ is the covariance due to unmodeled point sources and $\bR^G$ is the covariance of Galactic emission. We now describe our model of the covariances for these two emission sources. 
\subsubsection{Unmodeled Point Sources}\label{sssec:PointSources}
With the MWA, point sources are completely sampled down to $\sMin \approx 50-80$\,mJy \citep{Caroll:2016,HurleyWalker:2016,Line:2016} and on LOFAR down to the $\sMin \approx 0.1$\,mJy level \citep{Williams:2016} within the primary beam. We represent these sources with an achromatic version of the model from \citet{Liu:2011}. At these faint fluxes, the sources are isotropically distributed and the number of sources with fluxes between $S$ and $S+dS$ within an infinitesimal solid angle $d\Omega$ is well described by a random Poisson process with a power law mean \citep{DiMatteo:2002}. 
\begin{equation}
\frac{d^2N}{dS d\Omega} =k \begin{cases} \left( \frac{S}{S_*} \right)^{-\gamma_1} & S\le S_*\\
\left( \frac{S}{S_*} \right)^{-\gamma_2} & S > S_*
\end{cases},
\end{equation}
where $k=4000$Jy$^{-1}$sr$^{-1}$, $\gamma_1=1.75$, $\gamma_2=2.5$, and $S_*=0.88$\,Jy. %The average intensity from unmodeled point sources on the sky, in direction $bs$ is
%\begin{align}
%\langle I_r(\bs) \rangle = \int_0^{S_{min}} dS %\frac{d^2N}{dS d\Omega} S \equiv \muR.
%\end{align}

 Consider a visibility, $v_\alpha$, formed by antennas $i$ and $j$, that are separated by baseline $\bb_\alpha$. The covariance between two baselines $v_\alpha(\freq)$ and $v_\beta(\freq')$ at two frequencies, $\freq$ and $\freq'$, assuming un-clustered and flat-spectrum sources is
\begin{align}\label{eq:VisCovariancePoints}
\cR^P_{\alpha\beta}(\freq,\freq') &= \int_0^{\sMin} dS \frac{dN}{dS d \Omega} S^2 \int d\Omega |A(\bs)|^2 e^{-2 \pi i \bs \cdot (\bb_\alpha\freq-\bb_\beta \freq')/c} \nonumber \\
&= \varR \int d\Omega |A(\bs)|^2 e^{-2 \pi i \bs \cdot(\bb_\alpha \freq -\bb_\beta \freq' ) /c},
\end{align}
where $A(\bs)$ is the primary beam of each antenna which we assume are identical.
The Fourier convolution theorem tells us that the last integral in equation~\ref{eq:VisCovariancePoints} is equal to the convolution of the Fourier transform of the beam with itself evaluated at $(\bb_\alpha \freq/c - \bb_\beta \freq'/c)$ in the $uv$ plane. This quantity falls to zero when $|\bb_\alpha \freq/c - \bb_\beta \freq' /c|$ is larger than the diameter of the antenna aperture. Thus as long as two baselines are separated by a distance greater then the antenna aperture diameter, $R^P_{\alpha \ne \beta} \approx 0$. We may therefore ignore off diagonal terms in the residual covariance matrix for minimally redundant arrays. It turns out that the diagonal covariance assumption gives similar results, even for maximally redundant arrays (see Appendix~\ref{app:Redundancy}).

\subsubsection{Diffuse Galactic Emission}\label{sssec:Diffuse}
Diffuse Galactic emission is correlated on large angular scales. We may construct a simple model of this emission using the same steps we used to obtain equation~\ref{eq:VisCovariance} and assuming that $uv$ power spectrum of the diffuse emission does not evolve significantly over an antenna footprint. Under these assumptions, one can show that the covariance between two visibilities from diffuse emission is 
\begin{align}\label{eq:VisCovarianceDiffuse}
\cR^G_{\alpha \beta}(\freq,\freq') & \approx P_G(\bb_\alpha/\lambda_0) \int d\Omega |A(\bs)|^2 e^{-2 \pi i \bs \cdot( \bb_\alpha  \freq - \bb_\beta \freq' ) /c}  
\end{align}
where $P_G(\bu)$ is the power spectrum of diffuse galactic emission in the $uv$ plane and $\lambda_0$ is the wavelength of the center of the interferometer's band. To model $P_G(\bu)$ we use an empirical power law fit to the two dimensional power spectrum of a desourced and destriped \citep{Remazeilles:2015} Galactic emission map \citep{Haslam:1982} centered at RA=$60^\circ$, DEC=$-30^\circ$ and scaled from 408 to 150\,MHz using a frequency power law with a spectral index of $-0.6$ \citep{Rogers:2008,Fixsen:2011}\footnote{This power law is for spectral radiance. For brightness temperature, the spectral index is $-2.6$.}. We find that the angular power spectrum of galactic emission at 150\,MHz is well modeled by a power law in $u=|\bb|/\lambda_0$, $P_G(\bu) = 6 \times 10^{11} u^{-5.7}$Jy$^2$Sr$^{-1}$. Throughout this paper, we will assume that the model used for calibration completely ignores diffuse emission so that all of $R_{\alpha\beta}^G$ is included in the co-variance of residual visibilities.

\subsection{Frequency Domain Calibration Errors}\label{ssec:Linearized}

So far we have a model of the discrepancies between true and modeled interferometer visibilities. Given this model, what are the statistics of the errors in our frequency dependent gain solutions? Our goal in this subsection is to derive the covariances of errors in gain parameters in terms of the covariances of the visibility residuals discussed in \S~\ref{ssec:Sources}. 

We will start by writing down the system of equations that calibration algorithms attempt to solve, and, following W92 and L10, we will reduce this system to a set of linear equations that are valid in the regime of small calibration errors which is the case for errors generated by the faintest sources on the sky. This approximation holds when the gains are nearly correct after large gain variations are removed by a first iteration of calibration using a reasonably accurate calibration catalog. Writing down these systems in matrix form, the covariances of the least-squares solutions for these linear systems are readily obtained in the same manner as W92 and L10.

We start by writing down the equations that calibration must solve. In line with the notation of L10, we parameterize the small gain and phase of the $i^{th}$ antenna after rough calibration as the exponent of a complex number,
\begin{equation}
g_i(\freq) = e^{\eta_i(\freq) + i \phi_i(\freq)} \approx 1 + \eta_i(\freq) + i \phi_i(\freq),
\end{equation}
where $\eta_i$ is the amplitude of the gain and $\phi_i$ is the phase.
In calibration, one attempts to solve the set of equations
\begin{equation}\label{eq:Calibrate}
g_i(\freq) g_j^*(\freq) v_{ij}^{\text{true}}(\freq) = v_{ij}^{\text{meas}}(\freq)
\end{equation}
where $v_{ij}^{\text{meas}}(\freq)$ is the measured visibility.
 If we divide by $y_{ij}$ on both sides, we have
\begin{equation}\label{eq:CalibrateDivide}
g_i g^*_j  \left(1 + \frac{c_{ij}}{y_{ij}} \right)=\frac{v_{ij}^{\text{meas}}}{y_{ij}}.
\end{equation}
Recall that $c_{ij}$ represents the sum of unmodeled components of a visibility (equation~\ref{eq:Components}) while $y_{ij}$ represents the modeled component. 
For analytic tractability, we will linearize these equations by working to first order in $c_{ij}/y_{ij}$, $\eta_i$, and $\phi_i$. With this approximation, equation~\ref{eq:CalibrateDivide} becomes
\begin{align}
\frac{ v_{ij}^{\text{meas}}}{y_{ij}} & \approx (1 + \eta_i + i \phi_i )(1 +\eta_j - i \phi_j)\left(1 + \frac{c_{ij}}{y_{ij}}\right) \\
& \approx 1 + \eta_i + \eta_j + i\phi_i - i \phi_j +\frac{c_{ij}}{y_{ij}} 
\end{align}
Separating the real and imaginary parts gives us two systems of linear equations;
\begin{equation}\label{eq:LinearizedAmps}
\eta_i+\eta_j + \re \left(\frac{c_{ij}}{y_{ij}}\right) \approx \re \frac{v_{ij}^{\text{meas}}}{y_{ij}} - 1 \equiv \measEta_{ij}
\end{equation}
and 
\begin{equation}\label{eq:LinearizedPhase}
\phi_i - \phi_j  + \im \left(\frac{c_{ij}}{y_{ij}}\right) \approx \im \frac{v_{ij}^{\text{meas}}}{y_{ij}} \equiv \measPhi_{ij}.
\end{equation}
Since residual foregrounds may be described statistically, we treat $c_{ij}$ as a noise term in the same way that thermal noise is treated in L10. Unlike thermal noise, which is typically uncorrelated in frequency and ideally has the same variance across baselines, the correlation properties of modeling noise are those of the unmodeled sources discussed in \S~\ref{ssec:Sources}. We can write the system of equations given by equation~\ref{eq:LinearizedAmps} in matrix form\footnote{The system of equations used here only attempts to solve for the gains. In redundant calibration, the number of unique true visibilities is reduced to a point where one can also solve for them as-well. This leads to different forms for the matrix equations (see W92 and L10 for examples).},
\begin{equation}
\begin{pmatrix} \measEta_{12} \\ 
				\measEta_{23} \\
 				\vdots \\
 				 \measEta_{N-1N} \\ 
 				 \measEta_{13} \\ 
 				 \vdots \\ 
 				 \measEta_{N-2N} \\ 
 				 \vdots \\ 
 				 \measEta_{1N} 
 				 \end{pmatrix} 
 = \begin{pmatrix} 1 & 1 & 0 & \hdots & 0 \\ 
 				   0 & 1 & 1 & \hdots & 0 \\
 				   \vdots & \vdots & \vdots & \ddots & \vdots  \\
 				   0 & 0 & 0 & \hdots & 1 \\
 				   1 & 0 & 1 & \hdots & 0 \\
 					\vdots & \vdots & \vdots & \ddots & \vdots  \\
 					1 & 0 & 0 & \hdots & 0 \\
 					\vdots & \vdots & \vdots & \ddots & \vdots  \\
 					1 & 0 & 0 & \hdots & 1 
 				   
 \end{pmatrix}
 \begin{pmatrix}
 \eta_1 \\
 \eta_2 \\
 \eta_3 \\
 \vdots \\
 \eta_N
 \end{pmatrix} +\re\left(\frac{\bf c}{\bf y} \right),%+ \begin{pmatrix}
 % \re n_{12} \\
 % \re n_{23} \\
 % \vdots \\
 % \re n_{N-1N}\\
 %\re n_{13} \\
 %\vdots \\
 %\re n_{n-2N} \\
 %\vdots \\
 %\re n_{1N}
 %\end{pmatrix}
\end{equation}
which which we write more compactly as $\bmeasEta = \bA \bEta + \re( {\bf c}/{\bf y})$.
The same can be done for equation~\ref{eq:LinearizedPhase}.
\begin{equation}
\begin{pmatrix} \measPhi_{12} \\ 
				\measPhi_{23} \\
 				\vdots \\
 				 \measPhi_{N-1N} \\ 
 				 \measPhi_{13} \\ 
 				 \vdots \\ 
 				 \measPhi_{N-2N} \\ 
 				 \vdots \\ 
 				 \measPhi_{1N} \\
 				 0
 				 \end{pmatrix} 
 = \begin{pmatrix} 1 & -1 &  \hdots & 0 \\ 
 				   0 & 1  & \hdots & 0 \\
 				   \vdots & \vdots & \ddots  & \vdots \\
 				   0 & 0 &  \hdots & -1 \\
 				   1 & 0 & \hdots & 0 \\
 					\vdots & \vdots & \ddots & \vdots  \\
 					1 & 0 & \hdots & 0 \\
 					\vdots & \vdots & \ddots  & \vdots \\
 					1 & 0 & \hdots & -1 \\
 					1 & 1 & \hdots & 1 
 \end{pmatrix}
 \begin{pmatrix}
 \phi_1 \\
 \phi_2 \\
 \vdots \\
 \phi_N
 \end{pmatrix} + \im \left( \frac{\bf c}{\bf y} \right),%+ \begin{pmatrix}
  %\im n_{12} \\
  %\im n_{23} \\
  %\vdots \\
  %\im n_{n-1N}\\
 %\im n_{13} \\
 %\vdots \\
 %\im n_{n-2N} \\
 %\vdots \\
 %\im n_{1N} \\
 %0
 %\end{pmatrix}
\end{equation}
where the last row in the matrix arises from imposing the constraint (L10) that $\sum_j \phi_j = 0$\footnote{The arbitrary phase reference is often set in sky-based calibration by defining the phases as the differences between each antenna phase and that of an arbitrarily chosen reference antenna. This constraint can be written as, $\phi_{\text{ref}}=0$ and would modify the last row of $\bB$ to be zero except for the index of the reference antenna (rather than all ones as we have written it). While choosing the reference antenna form of the phase constraint affects the details of some of the expressions in this paper, it results in the same scaling relationships and has a negligible effect on quantitative results.} We write the imaginary equation as $\bmeasPhi = \bB \bphi + \im( {\bf c }/{\bf y} )$. 

Given a model and measurements of $\bmeasPhi$ and $\bmeasEta$, a least squares estimator that applies weights of $\bW$ to each measurement will arrive at solutions for $\bEta$ and $\bphi$ given by
\begin{equation}\label{eq:EtaEstimate}
\bdwEta = (\bAt \bW \bA)^{-1} \bAt \bW \bmeasEta  \equiv \bLm \bmeasEta
\end{equation}
and 
\begin{equation}\label{eq:PhiEstimate}
\bdwPhi =(\bBt \bW \bB)^{-1} \bBt \bW \bmeasPhi \equiv \bPm \bmeasPhi.
\end{equation}
We emphasize that $\bdwEta$ and $\bdwPhi$ are estimates of the true values, $\bEta$ and $\bPhi$. The covariance of these estimates,
\begin{align}
\bcEta(\freq,\freq') = \langle \bdwEta(\freq) \bdwEta\trans(\freq') \rangle - \langle \bdwEta(\freq) \rangle \langle \bdwEta\trans(\freq') \rangle \\
\bcPhi(\freq,\freq') = \langle \bdwPhi(\freq) \bdwPhi\trans(\freq') \rangle - \langle \bdwPhi(\freq) \rangle \langle \bdwPhi\trans(\freq') \rangle,
\end{align}
is given by 
\begin{align}
\bcEta(\freq,\freq') &= \bLm \cov\left[ \re\left(\frac{\bc}{\by} \right),\re\left(\frac{\bc}{\by}  \right)^\intercal \right] \bLmt \label{eq:covEta}\\
\bcPhi(\freq,\freq')  & = \bPm \cov \left[ \im\left(\frac{\bc}{\by} \right),\im\left(\frac{\bc}{\by}  \right)^\intercal \right] \bPmt  \label{eq:covPhi}
\end{align} 
Thus, we have arrived at expressions for the covariances of errors in the gain parameters in terms of the covariances of the real and imaginary components of the unmodeled visibilities. Equations~\ref{eq:covEta} and \ref{eq:covPhi} show that the covariance of any given gain solution is the linear combination of the covariances of every visibility in the array. Thus, the application of a gain solution (derived from an incomplete sky model) to a short baseline introduces the fine-frequency errors from long baselines. Our next step is to determine the impact of this leakage on the power spectrum.

\subsection{The Impact of Gain Errors on the 21\,cm Power Spectrum}\label{ssec:PowerSpectrum}
We now propagate the frequency dependent errors in each gain solution into the delay power spectrum. Calibration gives us an estimate of the gains,
\begin{equation}
\widehat{g}_i = e^{\widehat{\eta}_i + i \widehat{\phi}_i} \approx 1 + \widehat{\eta}_i + \widehat{\phi}_i.
\end{equation} 
whose deviations from the true gains ($\eta_i$ and $\phi_i$) have covariances given by equations~\ref{eq:covEta} and \ref{eq:covPhi}. The corrected, model-subtracted visibilities obtained from calibration are given by
\begin{align}\label{eq:fMultiply}
V_{ij} &=\frac{g_i g_j^*}{\hg_i \hg_j^*}(y_{ij} +c_{ij})-y_{ij} \nonumber \\
 	&\approx (y_{ij}+ c_{ij}) \times \nonumber \\
 	& [1 + (\eta_i-\widehat{\eta}_i )+ (\eta_j-\widehat{\eta}_j )+i (\phi_i -\widehat{\phi}_i)- i( \phi_j-\widehat{\phi}_j)]-y_{ij}
\end{align}
 The delay transform \citep{Parsons:2012a} is a popular estimate of the power spectrum in which visibilities are Fourier transformed from frequency into delay. Delay can be mapped approximately to Fourier modes along the LoS while the $uv$ coordinates of the visibility can be mapped to Fourier modes perpindicular to the LoS. The delay-transform is given by,
\begin{equation}\
\ftV_{ij}(\tau)  = \int d\freq e^{2 \pi i \freq \tau} V_{ij}(\freq),
\end{equation}
which we can apply to the gain-corrected and foreground subtracted visibility in equation~\ref{eq:fMultiply}. Taking the delay transform of equation~\ref{eq:fMultiply} and setting $\bEta' \equiv \bdwEta - \bEta$ and $\bPhi' \equiv \bdwPhi - \bPhi$  we have,
\begin{align}\label{eq:tMultiply}
\ftV_{ij}(\tau) &\approx -y_{ij} \star \left(\feta_i '+ \feta_j' + i \fphi_i' - i \fphi_j '\right) \nonumber \\
&+ \fc_{ij} - \fc_{ij} \star\left(\feta_i' + \feta_j' + i \fphi_i' - \fphi_j' \right),
\end{align}
where $\star$ denotes a convolution in delay space. For the sake of analytic tractability, we will ignore the chromaticity of $y_\alpha$ and set all $y_\alpha=S_0$, essentially assuming that that the modeled visibilities are dominated by a single source near the phase center which exceeds the flux of all other sources by a factor of several. Even with chromatic $y_\alpha$, per-frequency inverse covariance weighting, which multiplies each $\alpha^{th}$ weight by $|y_\alpha|^2$ (L10) removes some of this structure. In Appendix~\ref{app:ModelApprox} we explore the impact of relaxing this assumption and find that our achromatic $y_{ij}$ model predicts the LoS wave numbers at which the modeling noise drops below the 21\,cm power spectrum to within $\approx 10\%$ of what we find with chromatic $y_{ij}s$ obtained from a realistic sky model. Still, this dramatic assumption limits the accuracy of our specific quantitative predictions and we are exploring its impact in full calibration simulations. 

 The cosmological 21\,cm power spectrum, $P({\bf k})$, is well approximated by the mean amplitude square of the delay-transformed visibility multiplied by linear factors given in \citet{Parsons:2012b}
\begin{equation}\label{eq:DelayPS}
P(\bk) \approx \left(\frac{c^2}{2 k_B^2\freq_0^2} \right)^2 \frac{X^2(\freq_0)Y(\freq_0)}{B_{pp}\Omega_{pp}} \langle |\ftV(\bu,\eta)|^2 \rangle
\end{equation}
where $\Omega_{pp} = \int d\Omega |A(\bs)|^2$ and $B_{pp} = \int df |B(\freq)|^2$ are respectively the integrals of the squares of the beam and bandpass, $\freq_0$ is the center frequency of the observation, $k_B$ is the Boltzmann constant, and $(X,Y)$ are multiplicative factors converting between native interferometry coordinates and comoving cosmological coordinates, $2 \pi (u,v,\eta) = (Xk_x, Xk_y, Yk_z)$. In line with \citet{Parsons:2012b}, $\eta$ is used to denote the Fourier dual to frequency at fixed $|\bu|$ and $\tau$ to denote the frequency Fourier transform of a visibility which integrates over a slanted line in $u$-$\nu$ space. While this slanted integral introduces non-negligible mode-mixing (namely the wedge), it is a decent approximation for the range of $\eta$s probed by current and next-generation experiments. 

Therefore, we can estimate the power spectrum from calibration-modeling errors by cross-multiplying $\ftV_\alpha$ with its complex conjugate. If we denote the expectation value
\begin{equation}
\langle \ftV_\alpha(\tau)\ftV_\beta^*(\tau') \rangle \equiv \mathsf{P}_{\alpha \beta}(\tau,\tau'),
\end{equation}
 the bias from visibility residuals, is equal to $\mathsf{P}_{\alpha \alpha}(\tau,\tau)$ multiplied by the constant prefactors in equation~\ref{eq:DelayPS}. While we only need $\mathsf{P}_{\alpha \alpha}(\tau,\tau)$ for the bias, we will need off diagonal terms to calculate the variances of binned and averaged power spectrum estimates. We first write down  $\mathsf{P}_{\alpha \beta}(\tau,\tau')$ to second order in $\bc/\by$ with baseline $\alpha$ formed from antennas $i$ and $j$ and baseline $\beta$ formed from antennas $m$ and $n$,
\begin{align}\label{eq:CovVis}
&\langle \ftV_\alpha(\tau) \ftV^*_\beta(\tau') \rangle \equiv \mathsf{P}_{\alpha \beta}\nonumber \\ 
\approx S_0^2&\left[ \vphantom{ \langle \fc_\alpha \fphi_n^{\prime *} \rangle}  \langle \feta_i' \feta_m^{\prime*} \rangle +  \langle \feta_i' \feta_n^{\prime*} \rangle +  \langle \feta_j' \feta_m^{\prime*} \rangle +  \langle \feta_j' \feta_n^{\prime*} \rangle \right. \nonumber \\
&+ \langle \fphi_i' \fphi_m^{\prime*} \rangle - \langle \fphi_i' \fphi_n^{\prime*} \rangle  -  \langle \fphi_j' \fphi_m^{\prime*} \rangle + \langle \fphi_j' \fphi_n^{\prime*} \rangle \nonumber \\
&-i \langle \feta_i'\fphi_m^{\prime *} \rangle+ i  \langle \feta_i' \fphi_n^{\prime *} \rangle - i   \langle \feta_j' \fphi_m^{\prime *} \rangle + i  \langle \feta_j' \fphi_n^{\prime *} \rangle  \nonumber \\
&- \left. i \langle \fphi_i' \feta_m^{\prime *} \rangle + i  \langle \fphi_i' \feta_n^{\prime *} \rangle -i  \langle \fphi_j' \feta_m^{\prime *} \rangle - i \langle \fphi_j' \feta_n^{\prime *} \rangle \right] \nonumber \\
& - S_0 \left[\vphantom{ \langle \fc_\alpha \fphi_n^{\prime *} \rangle} \langle \feta_i' \fc_\beta^* \rangle - \langle \feta_j' \fc_\beta^* \rangle - i  \langle \fphi_i \fc_\beta^* \rangle + i \langle \fphi_j \fc_\beta^* \rangle \right. \nonumber \\
& -  \left. \langle \fc_\alpha \feta_m^{\prime *} \rangle -  \langle \fc_\alpha \feta_n^{\prime *} \rangle + i  \langle \fc_\alpha \fphi_m^{\prime *} \rangle - i   \langle \fc_\alpha \fphi_n^{\prime *} \rangle \right] \nonumber \\
& + \langle \fc_\alpha \fc_\beta^* \rangle.
\end{align}
For the sake of space, we do not explicitly write $\tau$ or $\tau'$ in every term but understand that every complex conjugated term in each product is a function of $\tau'$ and every non-conjugated term is a function of $\tau$.  

Equation~\ref{eq:CovVis} involves six types of terms; those involving cross-multiples of $\feta'$, cross-multiples of $\fphi'$, cross-multiples of $\feta'$ and $\fphi'$, products between $\feta$ and $\fphi$ with $\fc$, and finally the covariances of the residuals themselves. In Appendix~\ref{app:Approximate} we obtain approximate expressions for each of the first five terms when the baseline separation is longer than the antenna diameter,
\begin{align}
 \langle \feta_i'(\tau) \feta_m^{\prime*}(\tau')  \rangle & \approx \int d \freq d \freq' e^{2 \pi i (\freq \tau - \freq' \tau')}[\bcEta(\freq,\freq')]_{im} \nonumber \\
 & \equiv \frac{S_0^{-2}}{2} [\bLm\bftC(\tau,\tau')\bLmt]_{im} \nonumber \\
 & = \frac{S_0^{-2}}{2} \Lm_{i \gamma} \Lmt_{\delta m} \ftC^{\gamma \delta}(\tau,\tau')\label{eq:Simplification1} \\
 \langle \fphi_i'(\tau) \fphi_m^{\prime*}(\tau')  \rangle &	\approx \int d \freq d \freq' e^{2 \pi i  (\freq \tau - \freq' \tau')}[\bcPhi(\freq,\freq')]_{im} \nonumber \\
 &= \frac{S_0^{-2}}{2} [ \bPm \bftC(\tau,\tau') \bPmt ]_{im} \nonumber \\
 & = \frac{S_0^{-2}}{2} \Pm_{i \gamma} \Pmt_{\delta m} \ftC^{\gamma \delta}(\tau,\tau')\label{eq:Simplification2}  \\
 \langle  \fc_\alpha(\tau) \feta_m^{\prime*}(\tau')  \rangle & \approx \frac{1}{2} \Lm_{m \gamma} \int d \freq d \freq' e^{2 \pi i  (\freq \tau - \freq' \tau')}  [\bC(\freq,\freq')]^\gamma{}_\alpha \nonumber \\
 & = \frac{1}{2} \Lm_{i \gamma} \ftC^\gamma{}_\alpha(\tau,\tau')\label{eq:Simplification3}  \\
 \langle \fc_\alpha(\tau) \fphi_m^{\prime*}(\tau')  \rangle & \approx \frac{i}{2} \Pm_{m \gamma}\int d \freq d \freq' e^{2 \pi i  (\freq \tau  - \freq' \tau')} [\bC(\freq,\freq')]^\gamma{}_\alpha \nonumber \\
 & = \frac{i}{2} \Pm_{m \gamma} \ftC^\gamma{}_\alpha(\tau,\tau')\label{eq:Simplification4}  \\
 \langle \feta'_i(\tau) \fphi_m^{\prime*}(\tau') \rangle & \approx 0,\label{eq:Simplification5} 
\end{align}
where we used Einstein-notation with repeated raised and lowered indices to denote summation and have defined $\bftC(\tau,\tau')$ as the delay-transform of the $\bC$ matrix. $\ftC_{\alpha \beta}(\tau,\tau') \equiv \int d \freq d \freq' e^{-2 \pi i ( \tau \freq - \tau' \freq')} \covC_{\alpha\beta}(\freq,\freq')$. We also denote the delay-transform of the $\bN$, $\bR$, and $\bS$ matrices in a similar way as $\bftN$, $\bftR$, and $\bftS$. The final term in equation~\ref{eq:CovVis} is simply the covariance matrix of the delay-transformed residual visibilities, $\ftC_{\alpha\beta}(\tau,\tau')$.
Using the above identities, we may write equation~\ref{eq:CovVis} as 
\begin{align}\label{eq:CovVisSimplified}
\mathsf{P}_{\alpha \beta}(\tau,\tau') & = \frac{1}{2} \Big[ \Lm_{i \gamma} \Lmt_{ \delta m} + \Lm_{i \gamma} \Lmt_{\delta n} + \Lm_{j \gamma}  \Lmt_{\delta m} + \Lm_{j \gamma} \Lmt_{ \delta n} + \nonumber \\
&  \Pm_{i \gamma} \Pmt_{ \delta m} - \Pm_{i \gamma} \Pmt_{\delta n} - \Pm_{j \gamma}  \Pmt_{\delta m} + \Pm_{j \gamma} \Pmt_{\delta n} \Big] \ftC^{\gamma \delta}(\tau,\tau') \nonumber \\
& - \frac{1}{2} \left( \Lm_{i \gamma} + \Lm_{j \gamma} - \Pm_{i \gamma} + \Pm_{j \gamma} \right) \ftC^\gamma{}_\beta(\tau,\tau') \nonumber \\
& - \frac{1}{2} \left( \Lm_{m \gamma} + \Lm_{n \gamma} + \Pm_{m \gamma} - \Pm_{n \gamma} \right)\ftC^\gamma{}_\alpha(\tau,\tau) \nonumber \\
& +\ftC_{\alpha \beta}(\tau,\tau'),
\end{align}
 The power-spectrum bias in delay-transform estimates is given by $\mathsf{P}_{\alpha \alpha}(\tau,\tau)$  ($i=m$ and $j=n$),
\begin{align}\label{eq:VisibilityCovApprox}
\mathsf{P}_{\alpha \alpha}(\tau,\tau) &= \frac{1}{2} ( \Lm_{i \gamma} \Lmt_{\delta i} + 2 \Lm_{i \gamma} \Lmt_{\delta j} + \Lm_{j \gamma}\Lmt_{\delta j} )\ftC^{\gamma\delta}(\tau,\tau) \nonumber \\
& + \frac{1}{2} ( \Pm_{i \gamma} \Pmt_{\delta i} - 2 \Pm_{i \gamma} \Pmt_{\delta j} + \Pm_{j \gamma} \Pmt_{\delta j} )\ftC^{\gamma \delta}(\tau,\tau) \nonumber \\
& - (\Lm_{i \gamma} + \Lm_{j \gamma})\ftC^\gamma{}_\alpha(\tau,\tau)+ \ftC_{\alpha \alpha}(\tau,\tau).
\end{align}
 Equations~\ref{eq:VisibilityCovApprox} and \ref{eq:CovVisSimplified} show how calibration leaks unmodeled structure in every visibility, including the highly chromatic ones, into the power spectrum of otherwise smooth short baselines. The last term in equation~\ref{eq:VisibilityCovApprox} is the power spectrum of unmodeled foregrounds, noise, and the signal itself.  Recall that since foregrounds are naturally contained within the horizon delay of $b_\alpha$, it does not contribute power into the EoR window. The sums in the first two lines, on the other hand, mix the chromaticity of foregrounds on all baselines into the delay power spectrum of the $\alpha^{th}$ visibility. Baselines that are longer than $b_\alpha$ contribute emission at delays below their individual horizon-delays which are greater than the horizon delay of $b_\alpha$, allowing for contamination of the EoR window. 
 
Typically, an estimate of $\mathsf{P}_{\alpha \alpha}$ is obtained by cross-multiplying integration over independent time intervals and since noise within each interval is independent, $\cN_{\alpha\beta}(\freq,t;\freq',t')=0$ \citep{Dillon:2014} and we can ignore the thermal noise contribution to the bias given by equation~\ref{eq:CovVisSimplified}. However, a subtlety introduced by calibration errors is that if calibration solutions for the cross-multiplied visibilities are not derived from complementary time intervals, there will still exist a thermal noise bias arising from all but the last term in~\ref{eq:CovVisSimplified}. This is the case in \citet{Dillon:2015b}, \citet{EwallWice:2015a}, and \citet{Beardsley:2016b} where 0.5\,s time-steps are used for interleaving visibilities but 112\,s non-interleaved time-steps are used for calibration. This bias can also survive cross multiplying different redundant measurements of the same visibility as is done with PAPER \citep{Parsons:2014,Ali:2015}.
 
\subsection{Calibration Bias for a Simplified Model}\label{ssec:Individual}
 
How do the covariances between pairs of visibilities contribute to the final power spectrum? We showed in \S~\ref{ssec:Sources} that $\bR$, the covariance matrix of un-modeled foreground visibilities, is well approximated as a diagonal in minimally redundant arrays. The same is true for thermal noise which arises from independent fluctuations at each antenna and the 21\,cm signal. Thus, the first two lines of equation~\ref{eq:VisibilityCovApprox} are formed from the weighted sum of the variance of the $\nAnt(\nAnt-1)/2$ visibilities (where $\nAnt$ is the number of antenna elements) with $\gamma=\delta$ where the weight of each variance is given by $\Lm_{i \gamma} \Lmt_{\gamma j}$ and $\Pm_{i \gamma} \Pmt_{\gamma j}$. These values depend crucially on our choice of visibility weighting, $\bW$, but it is highly instructive to examine the case where $\bW$ is equal to the identity.  In Appendix~\ref{app:LambdaSums} we use  matrix algebra to that in the case of $\bW$ equal to the identity,
\begin{equation}\label{eq:LMatrix}
\Lm_{i \gamma} = \begin{cases} \frac{1}{\nAnt-1} & i \in \gamma \\ \frac{-1}{2(\nAnt-1)(\nAnt-2)}  &  i \not \in \gamma \end{cases}.
\end{equation}
and 
\begin{equation}\label{eq:PMatrix}
\Pm_{i \gamma} = \begin{cases} \frac{1}{\nAnt}  & \gamma = (i,\cdot) \\
- \frac{1}{\nAnt} & \gamma= (\cdot,i) \\ 0  &  i \not \in \gamma \end{cases},
\end{equation}
where we denote $\gamma=(i,\cdot)$ to denote any visibility with $i$ as the non-conjugated antenna and $\gamma=(\cdot,i)$ to be any baseline with $i$ as the conjugated antenna. Equation~\ref{eq:LMatrix} makes intuitive sense if we recall that $\Lm_{i\gamma}$ is the linear weight multiplied by each $\measEta_\gamma$ that is summed to form the $i^{th}$ gain solution. Inspecting equation~\ref{eq:LinearizedAmps}, we see that summing all $\nAnt-1$ $\Lm_{i \gamma}\measEta_\gamma$, that antenna $i$ participates in gives us
\begin{equation}
 \sum_{\gamma \ni i} \Lm_i{}^\gamma \measEta_\gamma = \eta_i + \frac{1}{\nAnt-1} \sum_{k \ne i} \eta_k
\end{equation}
 To remove the extra sum, and isolate $\eta_i$, we must subtract the sum all $\measEta$ that do not include the $i^{th}$ antenna, divided by $\nAnt-1$. For each $k\ne i$, there are $\nAnt-2$ baselines that involve $k$ but not $i$, so we must also divide each term by $\nAnt-2$. This gives us the weights for baselines not involving the $i^{th}$ antenna in equation~\ref{eq:LMatrix}. We can apply similar logic to equation~\ref{eq:PMatrix} by inspecting equations~\ref{eq:LinearizedPhase} and \ref{eq:PMatrix}. 

The weight with which the covariance between each pair of measurements contributes to the total covariance of $\bdwEta$ and $\bdwPhi$ is just the product of the weight with which each measurement is linearly summed.
\begin{equation}\label{eq:etaWeights}
\Lm_{i \gamma} \Lmt_{\delta j} = \begin{cases} \frac{1}{(\nAnt-1)^2}& i \in \gamma\text{ and }j \in \delta \\
\frac{-1}{(\nAnt-2)(\nAnt-1)^2}  & i \in \gamma \text{ and } j \not\in \delta \\
\frac{-1}{(\nAnt-2)(\nAnt-1)^2} & i \not\in \gamma \text{ and } j \in \delta \\
\frac{1}{(\nAnt-1)^2(\nAnt-2)^2} & i \not\in \gamma \text{ and } j \not\in \delta
\end{cases} 
\end{equation}
and
\begin{equation}\label{eq:phiWeights}
\Pm_{i \gamma} \Pmt_{\delta j} = \begin{cases} \frac{1}{\nAnt^2} & \gamma=(i,\cdot),\delta=(j,\cdot) \text{ or } \gamma=(\cdot,i), \delta=(\cdot,j) \\
-\frac{1}{\nAnt^2} & \gamma=(i,\cdot),\delta=(\cdot,j) \text{ or } \gamma=(\cdot,i), \delta=(j,\cdot) \\
0 & \text{ otherwise. }
\end{cases} 
\end{equation}

For non-redundant arrays, $\bC(\freq,\freq')$ is diagonal and we can focus on $\gamma=\delta$ terms. From equations~\ref{eq:etaWeights} and \ref{eq:phiWeights} we see that when $i=j$, each visibility variance is weighted by $\sim \nAnt^{-2}$ when $i \in \gamma$ and at most by $\nAnt^{-4}$ when $i \not \in \gamma$. Since there are $\sim \nAnt$ visibilities with antenna $i$ and $\sim \nAnt^{2}$ visibilities without antenna $i$, $i=j$ terms contributing to $\mathsf{P}_{\alpha \alpha}$ are given roughly by the average of visibility variances not involving $i$ divided by $\sim \nAnt^{-2}$ plus the average of visibility variances involving antenna $i$ divided by $\sim \nAnt$.
\begin{align}
\Lm_{i \gamma}\Lmt_{\delta i} \ftC^{\gamma\delta}(\tau,\tau) & \approx \frac{1}{\nAnt^2} \langle \ftC_{\delta\delta}(\tau,\tau) \rangle_{i \not \in \delta} + \frac{1}{\nAnt} \langle  \ftC_{\delta \delta}(\tau,\tau) \rangle_{i \in \delta } \nonumber \\
& \approx \frac{1}{\nAnt} \langle  \ftC_{\delta \delta}(\tau,\tau) \rangle_{i \in \delta} 
\end{align}
 where the $\langle \rangle_{i\in\delta}$ indicate an average over the set of baselines that include antenna $i$ and $\langle \rangle_{i \not \in \delta}$ denotes an average over baselines that are not formed using antenna $i$. The same equation holds for the $\Pm$ sums. Considering how elements of $\bC(\freq,\freq')$ scale with $\nAnt$, we can see the average of visibilities involving antenna $i$ dominate $\mathsf{P}_{\alpha \alpha}$ by a factor of $\nAnt$. 

For $i\ne j$, there is exactly one visibility that involves both antennas and will be weighted at most by $\nAnt^{-2}$. The $\sim \nAnt$ visibilities formed from $i$ XOR $j$ are weighted by $\nAnt^{-3}$ and the $\sim \nAnt^2$ visibilities that involve neither $i$ nor $j$ are weighted by $\nAnt^{-4}$. Thus all terms with $i\ne j$ in equation~\ref{eq:VisibilityCovApprox} give contributions on the order of the average of the visibility variances divided by $\nAnt^{2}$. 

It follows that if $\bC(\freq,\freq')$ is diagonal for all $\freq,\freq'$ and $\bW$ is equal to the identity, $i=j$ sums in equation~\ref{eq:VisibilityCovApprox} dominate by $\sim \nAnt$ and are well approximated by the average visibility variance involving antenna $i$ or $j$ divided by $\nAnt$. The overall level of foreground contamination from calibration errors therefore goes as $\nAnt^{-1}$ with the details of the extent in delay contamination depending on the antenna distribution and primary beam. Replacing each $i=j$ sum in equation~\ref{eq:VisibilityCovApprox} with an average over visibility covariances involving $i$ and $j$ and ignoring $i \ne j$ sums we arrive at an approximate formula that can be readily used to estimate $\mathsf{P}_{\alpha \alpha}$. 

\begin{align}\label{eq:CovVisSuperSimple}
\mathsf{P}_{\alpha \alpha}(\tau,\tau) &\approx \frac{1}{ \nAnt} \left[  \langle \ftC_{\delta \delta}(\tau,\tau) \rangle_{i \in \delta} + \langle \ftC_{\delta \delta}(\tau,\tau) \rangle_{j \in \delta}  \right]\nonumber \\
& + \ftC_{\alpha \alpha}(\tau,\tau) 
\end{align}
The two assumptions going into this formula are that for each $\freq$ and $\freq'$, $\bC(\freq,\freq')$ is diagonal (minimal redundancy) and that $\bW$ is set equal to unity. Equation~\ref{eq:CovVisSuperSimple} illustrates how the bias of a power spectrum estimate depends on both the covariance of the individual baseline (the second term) and the covariances of the baselines that share common antennas. In other words, the measurement of the power spectrum for a given baseline and delay depends on both on the residual foregrounds, noise, and signal at that baseline and delay and, suppressed by a factor of $\nAnt$, that of all other baselines at that delay that share an antenna with it.

\section{Modeling Noise in Existing Arrays}\label{sec:Results}
 Having developed our formalism in \S~\ref{sec:Formalism}, we may obtain order-of-magnitude estimates for the visibility modeling noise using equation~\ref{eq:CovVisSimplified} for four existing or planned arrays; LOFAR, MWA, HERA, and the re-baselined SKA-LOW. We discuss our models of each instrument (\S~\ref{ssec:Instruments}). We then determine the level of modeling noise in (\S~\ref{ssec:NoiseResultsExisting}) and the impact of beam modeling errors (\S~\ref{ssec:BeamErrors}). Equation~\ref{eq:CovVisSuperSimple} can be used to provide us with some intuition for how the properties of the noise scales with those of the array and catalog. In \S~\ref{ssec:Scaling}, we discuss these scalings and to what extent they may be used to reduce the amplitude of modeling noise. In each simulation, we assume that the foreground model, used for calibration and subtraction, contains point sources modeled down to some minimal flux level $S_\text{min}$ and that the true sky contains both the diffuse emission and all point sources. 
 \subsection{Instrumental Models}\label{ssec:Instruments}

 For all arrays, we consider an Airy beam for an aperture with diameter $\dAnt$, 
\begin{equation}\label{eq:Airy}
A(\bs) = A(\theta)= \left(2 \frac{J_1(\pi \dAnt \cos \theta /\lambda_0)}{\pi \dAnt \cos \theta \lambda_0} \right)^2,
\end{equation}
 where $\theta$ is the arc length from the beam pointing centre. An Airy beam has the virtue of a simple analytic expression that, unlike a Gaussian beam, exhibits realistic side-lobe structure which in turn affects foreground contamination near the edge of the wedge \citep{Thyagarajan:2015a,Pober:2016}.

Strictly speaking, the primary beam evolves with frequency; however, we find in numerical calculations that allowing for this variation has a negligible impact on our results. In order to expedite the computation of calibration noise, especially for the arrays with large numbers of antennas, such as HERA and the SKA-1, we also assume that $\bR$ is diagonal. This is clearly not the case for the highly redundant HERA layout but we find  (Appendix~\ref{app:Redundancy}) that this only impacts the amplitude of the modeling noise by a factor of order unity and has a negligible impact on which modes are contaminated. In all arrays, we assume that every baseline is given equal weighting of unity. Note that for the SKA and LOFAR, we do not explicitly include outrigger antennas in our model of calibration. Our models for each individual instrument are as follows.

\begin{itemize}
\item {\bf The Murchison Widefield Array} For the MWA, we use the 128 tile layout described in  \citet{Beardsley:2012} and \citet{Tingay:2013a}. Antennas are modeled as $4$\,m diameter circular apertures. We assume a flux limit of $86$\,mJy, which is the limit for the array's naturally weighted point spread function at $150$\,MHz and similar to limits obtained in \citet{Carroll:2016}. Other analyses have obtained complete samples down to $35-50$\,mJy \citep{HurleyWalker:2016} but this order unity change in $\sMin$ does not significantly impact the modeling noise level which scales as $\sim \sMin^{1.25}$ (see \S~\ref{ssec:Scaling}). The deeper TIFR GMRT Sky Survey (TGSS) covers a significant portion of the MWA's field of view and is complete down to $10$\,mJy \citep{Intema:2016}. We therefore also consider an optimistic scenario where a deep TGSS catalog is used to calibrate the instrument. 

\item {\bf The Low Frequency Array}
We model LOFAR as the 48-element high band core described in \citet{VanHaarlem:2013}, with 30\,m diameter circular stations. The confusion limit for the naturally weighted core is $\approx 35$\,mJy at $150$\,MHz. However, the use of LOFAR's extended baselines measures source catalogs that are complete down to $\sMin \approx 0.1$\,mJy \citep{Williams:2016}. While the \citep{Williams:2016} survey is over a $\approx 4^\circ$ field of view, the catalog we consider here covers the entire sky. Such a catalog would involve accurately characterizing $\sim 27$\,million sources and may not happen before the SKA but we consider it as a very optimistic bracket on LOFAR's performance. 

\item{\bf The Hydrogen Epoch of Reionization Array}
For our model of HERA, we use the 331 element hexagonally packed core described in \citet{Pober:2014}. Each element is modeled as a 14\,m diameter circular aperture. HERA is designed to be calibrated redundantly  \citep{Dillon:2016}; hence, the power spectrum estimates it obtains will not directly be affected by the modeling errors we explore in this paper. We choose to include HERA in order to assess the performance of compact cores in sky-based calibration and to explore sky-based calibration as an alternative to redundant calibration. HERA's confusion limit is $\sMin \approx 11$\,Jy. However, the dec $\approx -30^\circ$ stripe that it will scan is also covered by the TGSS survey which is complete down to  $\approx 10$\,mJy. We therefore also consider an optimistic scenario in which the TGSS catalog is used for calibration. 
 %We also explore the performance of HERA in sky-based calibration. HERA is designed to be fully redundantly calibratable \citep{Dillon:2016} but it is useful to assess the performance of sky-based calibration as an alternative with potentially different systematics. Since redundant calibration does not rely as much on a model of the sky, there exists the possibility of this array outperforming any of the predictions in this paper.

\item {\bf The Square Kilometre Array}
We investigate the level of modeling noise in the SKA-1 Low design proposed in \citet{Dewdney:2013} but scaled down to half of the described collecting area due to recent rebaselining. The array consists of 497, 30\,m stations with a number density distributed as a Gaussian in radius where 75\,\% of antennas fall within 1\,km of the center, corresponding to a standard deviation of $\sigmaAnt\sim600\,$m. The confusion limit of the SKA's core is $\approx 27$\,mJy, however the inclusion of extended baselines out to $\approx 100$\,km will bring the confusion limit at $150$\,MHz to $\approx 0.1$\,mJy \citep{Prandoni:2015} which we also consider as an optimistic case.

\end{itemize}

\subsection{Modeling Results}\label{ssec:NoiseResultsExisting}

\begin{figure*}
\includegraphics[width=\textwidth]{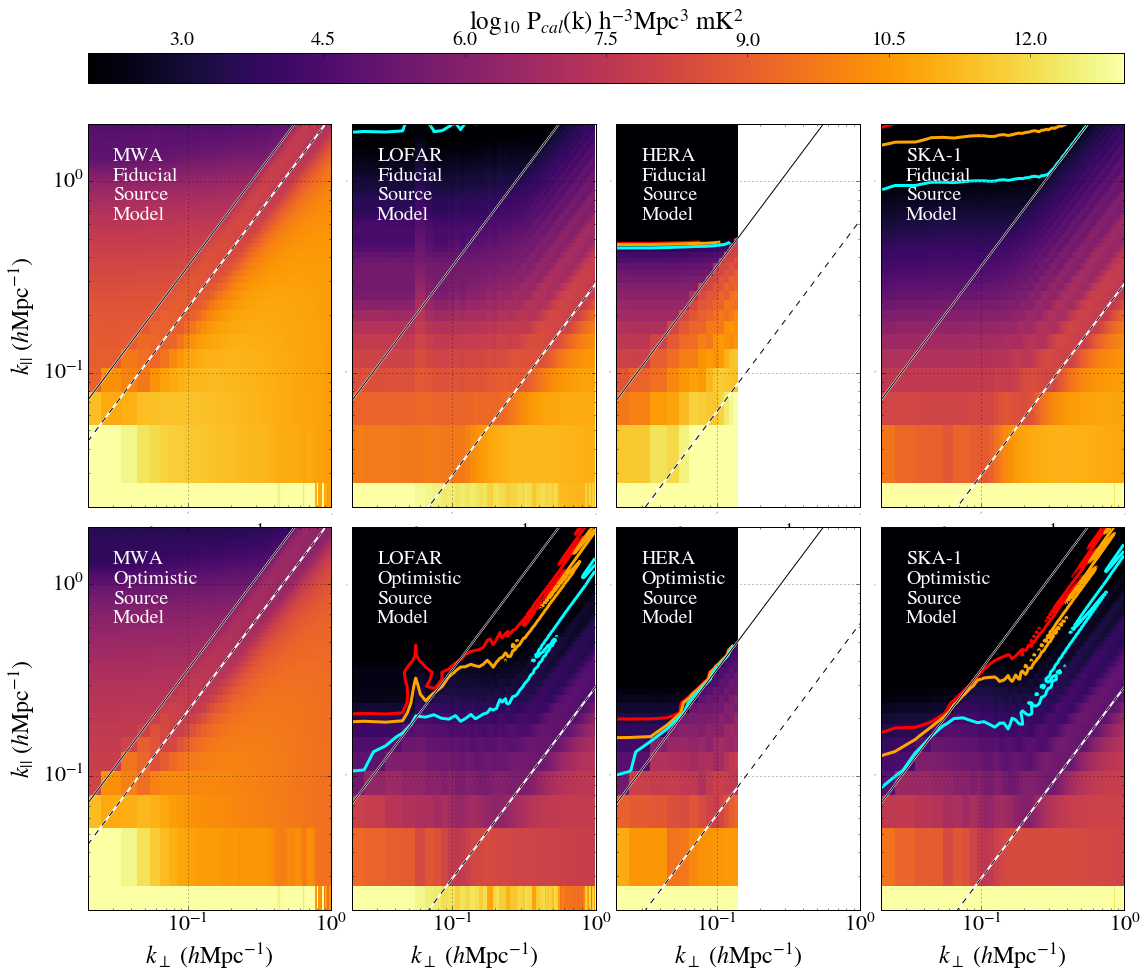}
\caption{Top: The power spectrum residuals computed using equation~\ref{eq:VisibilityCovApprox} for the MWA, LOFAR, HERA, and the SKA-1 LOW designs with sources modeled down to the array confusion limit. Unmodeled foregrounds are contained within the wedge, which is demarked by dashed black lines at the first primary beam null and solid black lines at the horizon. The calibration errors introduced by these foregrounds bleed out of the wedge into the EoR window. The narrower central lobes (larger stations) employed by LOFAR and the SKA help to significantly reduce the leakage at large $k_\parallel$ that exists for the MWA.  Contours where the signal, from a {\tt 21cmFAST} simulation, is equal to unity, five, and ten times the calibration noise are indicated by cyan, orange, and red lines respectively. Bottom: The same as the top for optimistic scenarios. The optimistic scenario for LOFAR and the SKA involves complete modeling of point sources down to $0.1$\,mJy using additional long baselines. For HERA and the MWA, the optimistic scenario assumes that the $10$\,mJy source catalog from the TGSS is used for calibration. If long baselines can faithfully model the sources down to $0.1$\,mJy, modeling noise does not appear to limit LOFAR and the SKA. Sky-based calibration with HERA is improved significantly by using a deep source catalog from a complementary array. The vertical stripe in the LOFAR figure at $\kperp \sim 0.6 h$Mpc$^{-1}$ arises from the arrangement of the HBA antennas in short spaced pairs so that even the outrigger antennas, which are heavily contaminated, participate in a single short baseline.}
\label{fig:CalWedges}
\end{figure*}

In Fig.~\ref{fig:CalWedges}, we plot cylindrically binned and averaged delay-transform power spectra of residual visibilities from unmodeled foregrounds, calculated using equation~\ref{eq:VisibilityCovApprox} for the MWA, LOFAR, HERA, and SKA-1. We explore two different $\sMin$ values for each array. As we might expect, the majority of residual power is contained within the wedge, arising from the last term in equation~\ref{eq:VisibilityCovApprox}. This term is the power spectrum of the unmodeled residual sources and would exist in the absence of calibration errors. For the MWA, which has a smaller aperture, hence a wider primary beam, the wedge of unmodeled sources extends to larger $k_\parallel$ values. Beyond the wedge extends the power spectrum of calibration errors which exist at the level of $10^6-10^8$\,$h^3$Mpc$^{-3}$\,mK$^2$; one to two orders of magnitude greater than the 21\,cm signal. For the MWA, the  level of contamination inside of the EoR window is within an order of magnitude of the simulated errors encountered in B16 who consider a calibration catalog that is incomplete to a similar depth. It is apparent that the calibration errors experience a sharp cutoff at the $k_\parallel$ corresponding to the delay of the edge of the main lobe on the longest baselines of the array. A vertical stripe of additional contamination appears in the LOFAR plot at $\kperp \approx 0.6 h$Mpc$^{-1}$ which corresponds to separation scale for the HBA antenna pairs. Since even the longest outriggers participate in a short baseline with this length, more significant contamination is introduced at the corresponding Fourier mode. 

We also estimate the region of $k$-space in which the 21\,cm signal will be accessible by computing the ratio between the 2d power spectrum of residual visibilities and a representative signal computed using {\tt 21cmFAST}\footnote{\url{http://homepage.sns.it/mesinger/DexM___21cmFAST.html}} \citep{Mesinger:2007,Mesinger:2011}. The reionization parameters are set to $T_{\text{vir}}^{\text{min}} = 2 \times 10^4 K$, $\zeta = 20$, and $\text{R}_{\text{mfp}} = 15$\,Mpc, yielding a redshift of 50\% reionization of $\approx 8.5$. For fiducial catalog limits, we see that the entire EoR window is unusable for LOFAR and the MWA while the SKA is only able to detect the signal at large $k \gtrsim 0.4$\,$h$Mpc$^{-1}$. If LOFAR and the SKA use their extended baselines to obtain deep source catalogs and calibrate on these catalogs {\it with only their core antennas}, they will be able to isolate modeling errors to be contained primarily within the wedge. By calibrating on a deep 10\,mJy catalog such as the TGSS, HERA can rely on traditional sky-based calibration as a potential alternative to its primary redundant strategy. 

\subsection{The Impact of Primary Beam Modeling Errors}\label{ssec:BeamErrors}
So far, we have assumed that the antenna primary beam is known perfectly. Here we examine the impact of an imperfectly modeled beam on calibration noise. It is worth noting that while we only examine the impact of beam errors on calibration, errors in beam modeling can affect other aspects of the analysis (for example the power spectrum normalization equation in equation~\ref{eq:DelayPS}). Recent in-situ measurements with Orbcomm satellites \citep{Neben:2015,Neben:2016} indicate that electromagnetic modeling of instrumental primary beams may only be accurate to the $1\%$ level within the central lobe and only to the $10\,\%$ level within the side-lobes. Even if a complete model of the sky exists, systematic errors in the apparent flux of these sources will cause calibration errors similar to those encountered in \S~\ref{ssec:Instruments}. We describe beam-modeling errors as an angle dependent function, $D(\bs)$ that is added to the known component of the beam $B(\bs)$. 
\begin{equation}
A(\bs) = B(\bs) + D(\bs).
\end{equation}

For the purposes of this section only, we take the optimistic case that we have a perfect external catalog and that all calibration modeling error comes from an incorrect model of the primary beam. A true visibility in the presence of these errors is 
\begin{align}
v_\alpha^{\text{true}} &= \int d \Omega [B(\bs) + D(\bs)]I(\bs)e^{-2 \pi i \bb_\alpha \freq/c} \nonumber \\
& = y_\alpha + \int d\Omega D(\bs) I(\bs)e^{-2 \pi i \bb_\alpha \freq/c}.
\end{align}
Our new calibration residual, $r_\alpha$ takes on the form:
\begin{align}
&r_\alpha \to \int d\Omega D(\bs) I(\bs)e^{-2 \pi i \bb_\alpha \freq/c}.
\end{align}
This leads to a new form of $R_{\alpha \beta}$ as well:
\begin{align}
&R_{\alpha \beta} \to \int_{0}^{\sMax} dS \frac{d^2N}{dS d\Omega}S^2\int d \Omega |D(\bs)|^2 e^{-2 \pi i (b_\alpha \freq/c - b_\beta \freq'/c)}, 
\end{align}
where $\sMax$ is the  flux of the highest flux source in the field of view which is obtained by setting the number of sources with intrinsic flux greater than $\sMax$ equal to unity,
\begin{equation}
\sMax = S_* \left[k \int d\Omega A(\bs) \right]^{1/\gamma}.
\end{equation}
 Since the literature typically reports fractional errors in beam-modeling, we describe $D(\bs)$ as the true beam $A(\bs)$ multiplied by a fractional error function $D(\bs) = f(\bs) A(\bs)$ where we parameterize $f(\bs)$ as the following piecewise function, 
\begin{equation}\label{eq:BeamErrorModel}
f(\bs) = \begin{cases}A\left[1-(1-e_z) \exp(-\cos \theta^2/2 \sigma_e^2) \right]& |\cos \theta|<s_1 \\ A\left[ 1-(1-e_z)\exp(-s_1^2/2 \sigma_e^2) \right] & |\cos \theta|\ge s_1, \end{cases}
\end{equation}
where $A e_z$ is the fractional beam-modeling error at the pointing centre, $s_1$ is the angular distance of the pointing center to the first side-lobe and $A,\sigma_e$ may be adjusted to give different fractional modeling errors in the side-lobes. This function allows us to assign an arbitrary modeling uncertainty to the zenith and side-lobes. We compute the level of beam-modeling noise in the 21\,cm power spectrum for two different scenarios, one in which the beam is known to $1\,\%$ at zenith and $10\,\%$ in the side-lobes, which is consistent with the precision reported in \citet{Neben:2015}. We also consider a scenario in which an order of magnitude improvement in beam modeling has been achieved and the beam is known to $1\,\%$ in both the side-lobes and the main lobe which is the target precision for in-development drone experiments \citep{Jacobs:2016b}. We note that our model describes beam modeling errors that are completely correlated between antennas. It is possible that uncorrelated errors (which we might expect to arise from imperfections in the construction of each station) will integrate down differently from the modeling errors we consider here. %The results are shown in Fig.~\ref{fig:BeamErrors}.

\begin{figure*}
\includegraphics[width=\textwidth]{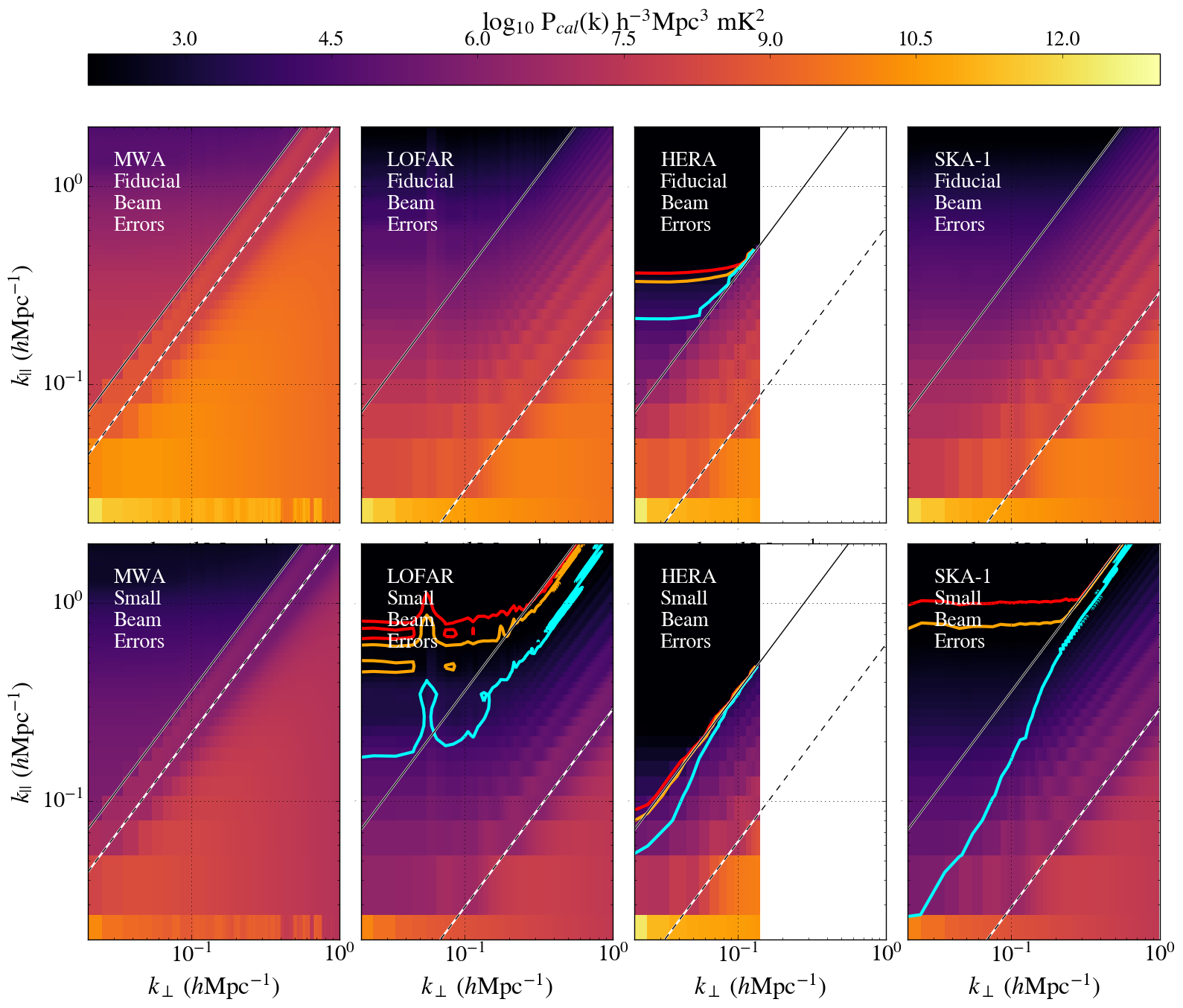}
\caption{Top: Same as Fig.~\ref{fig:CalWedges} except now we consider a perfect calibration catalog with calibration errors arising only from mismodeling the primary beam at the $1\%$ level in the main-lobe and the $10\%$ level in the side-lobes. With the exception of HERA, foreground residuals arising from primary beam modeling errors dominate the signal in the entire EoR window.  Bottom: the same as above, but with a uniform modeling accuracy of $1\%$. Much of the EoR window is still contaminated for LOFAR and the MWA while significant bias exists in much of the EoR window for the SKA.}
\label{fig:BeamErrors}
\end{figure*}

 Plotting the resulting residual power spectra from equation~\ref{eq:CovVisSimplified} in Fig.~\ref{fig:BeamErrors}, we see that with the current precision of primary beam models, the calibration noise masks the power spectrum across all of $k$-space for the MWA, LOFAR, and the SKA. Even with an order of magnitude improvement in our modeling, a significant foreground bias $\gtrsim 20\%$ of the signal amplitude will be present in measurements by LOFAR and SKA-1. Thus, even under the optimistic foreground modeling scenarios considered above, foreground errors will still contaminate the EoR window unless significant improvements in beam modeling are made. HERA's compact layout limits the impact of beam-errors to small delays so that a significant portion of the EoR window is accessible even in the fiducial beam-modeling scenario.

\subsection{The Dependence of Modeling Noise on Array and Catalog Properties}\label{ssec:Scaling}
We can use the equations developed in \S~\ref{ssec:PowerSpectrum} to determine the impact of array configuration and catalog depth on the power spectrum bias $\mathsf{P}_{\alpha \alpha}$. Since we are interested in the contribution from modeling errors which, unlike thermal noise, do not average down with integration time, we will let $\bC = \bR$ in equations ~\ref{eq:CovVis}-\ref{eq:CovVisSimplified}. From equation~\ref{eq:CovVisSuperSimple}, we list the effects of changing various parameters in the instrument design and source catalog in Table~\ref{tab:Params}. There are a number of adjustments in the array layout that can be made to reduce the amplitude of the errors. Several of these adjustments have multiple effects that work against each other. %In the following subsections we examine each of these strategies one by one. We explore the gains that can be made by improving the source model using observations on outrigger antennas or external arrays with higher resolution (\S~\ref{ssec:FluxLevels}), averaging calibration solutions from multiple independent fields (\S~\ref{ssec:TimeAveraging}), adjusting the number of elements, and compactness of the antenna distribution (\S~\ref{ssec:Compactness}).

\begin{table*}
\begin{tabular}{|p{2.4in}|p{3.6in}|}
\hline\hline
\textbf{Strategy} & \textbf{Impact on Error $P(\bk)$}  \\ \hline\hline
Reduce $\sMin$. & Reduce amplitude as $\sMin^{3-\gamma}$. \\
\hline
%Average calibration over $N_p$ fields. & Reduce amplitude by $N_p^{-1}$.\\
%\hline
Reduce the standard deviation of antenna positions, $\sigmaAnt$. & (a) Increase amplitude as $\sigma_a^{2(3-\gamma)/(1-\gamma)}$.  \\
 & (b) Reduce maximum $\kparmin \sim \sigma_a$ of errors. \\
\hline
Increase $\nAnt$. & (a) Reduce amplitude as $\nAnt^{-1}$. \\
& (b) Decrease $\sMin$.$^\dagger$ \\ 
\hline
Increase aperture diameter, $\dAnt$. &  (a) Reduces $\kparmin \sim \dAnt^{-1}$. \\
 &  (b) Requires larger $\sigmaAnt$ leading to larger $\kparmin$. \\
\hline
\multicolumn{2}{c}{$^\dagger$ {\it Depends on the distribution of additional antennas. }}
\end{tabular}
\caption{Inspection of the equations in \S~\ref{sec:Formalism} yields a number of analytic and qualitative relationships between the properties of an array and modeling catalog.}
\label{tab:Params}
\end{table*}

\subsubsection{Catalog Depth} %\label{ssec:FluxLevels}
Ignoring diffuse emission, the power spectrum of modeling errors in equation~\ref{eq:CovVisSuperSimple} is proportional to $\sigma_r^2=\int_0^{\sMin} S^2 S^{-\gamma}dS \propto \sMin^{3-\gamma}$. With the power law index of $1.75$ for faint source populations, the noise level will scale as $\sMin^{1.25}$. Hence, clearing a contaminated region requires improvements in catalog depth on the same order of magnitude as the ratio of the bias to the expected signal.

\subsubsection{Time Averaging} 
If the instrumental gains are stable in time, modeling noise can be suppressed by averaging over LSTs. We investigate the level of supression that is possible for {\it non fringe-stopped} baselines using multi-field averaging by calculating the temporal coherence of the modeling noise over some time interval, $\Delta t$. After a time $\Delta t$ has passed, the primary beam of the instrument that had a gain of $A(\bs)$ towards the brightness field at $I(\bs)$ at time $t$ will now have a gain of $A(\bs)$ towards $I(\bs+\bDs)$ at time $t+\Delta t$, 
\begin{align}
\cR_{\alpha\alpha}(\nu,t;\nu,t+\Delta t)= &\left(\int d\Omega d\Omega' e^{-2 \pi i \freq \bb_\alpha \cdot (\bs-\bs')/c} \right) \times \nonumber \\
&  \cov[I_r(\bs),I_r^*(\bs'+\bDs)] \nonumber \\
=&  \left(\varR e^{-2 \pi i \freq \bb_\alpha \cdot \bDs/c}\right) \times \nonumber \\ 
&\int d\Omega A(\bs) A^*(\bs - \bDs).\label{eq:TempCoherence}
\end{align}
When $\bDs$ is larger than the extent of the beam on the sky, the integral in equation~\ref{eq:TempCoherence} is close to zero. Hence a baseline is temporally coherent with itself when $\bDs$ is small enough that its fields of view at the different times overlap. If the gains are stable over time, one can calibrate on multiple fields and reduce the power spectrum of calibration modeling errors by a factor of $N_p$, where $N_p$ is the number of non-overlapping pointings. More significant suppression can arise from the oscillating term in equation~\ref{eq:TempCoherence} which arises from our assumption that the sky has moved by $\bDs$ and would not appear in the covariance between non fringe-stopped baselines. Averaging this oscillatory term over multiple LSTs can potentially lead to a significant reduction in the amplitude of modeling noise and is the subject of future work. %For a rough estimate of the gain from calibrating on multiple fields, we assume that the covariance between different times is negligible once $\Delta \widehat{s}$ is equal to the first null of an airy beam with an aperture equal to the station diameter, $\Delta \widehat{s} \approx 3.8317 \lambda_0/(\pi \dAnt)$. For the $30$\,m aperture of LOFAR and the SKA, $\Delta \widehat{s} \approx 4^\circ$ and for the $4$\,m MWA aperture, $\Delta \widehat{s} \approx 35^\circ$. Thus, there are $\approx 10$ independent fields of view that the MWA can average calibrations over and $\approx 100$ that LOFAR and the SKA can average over. However, in order to harness this supression, the gains must be identical across all averaged pointings. 

\subsubsection{Array Configuration}
There are three primary ways of changing the array configuration to affect modeling errors.
\begin{enumerate}
\item{\bf Antenna Distribution:}
Reducing the length of baselines involved in calibration reduces the chromaticity of gain errors and thus the smallest Fourier mode, $\kparmin$, that is not dominated by modeling noise. On the other hand, the array point spread function (PSF), and hence the minimal flux that an array can model for self-calibration is also set by by its compactness. If the antennas are distributed as a Gaussian with standard deviation $\sigma_a$, than the naturally weighted PSF can be approximated by a Gaussian with standard deviation, $\sigma_p=\lambda_0/(2 \pi \sigma_a)$. \citet{Condon:1974} determine that the confusion limit of an array, $\sMin$, depends on the PSF as $\sigma_p^{2/(\gamma-1)}\propto \sigma_a^{2/(1-\gamma)}$. Since the amplitude of the calibration noise is proportional to $\sMin^{3-\gamma}$,  the overall normalization of calibration noise will scale with the standard deviation of the antenna distribution as $\sigma_a^{2(3-\gamma)/(1-\gamma)} \sim \sigma_a^{-3.33}$. At a glance, this is a very steep change in amplitude which might counteract the decrease in chromaticity. However, will find below that the impact of chromaticity is much more important.  
\item{\bf Antenna Count:}
Increasing the number of antennas will cause the amplitude of the modeling noise power spectrum to reduce as $\sim\nAnt$ but larger numbers of antennas will also force the array to be less compact, potentially increasing $\kparmin$ while driving down the confusion limit.
\item{\bf Antenna Size:}
 Increasing the size of each antenna reduces the primary beam width and hence the contamination from foregrounds at delays near the horizon but also drives up the minimal baseline size. 
\end{enumerate}

The scaling of the noise with the array characteristics listed above can be illuminated with some further simplifying assumptions. In particular, if all of the stations have Gaussian beams with angular standard deviations of $\sigmaBeam \approx \epsilon \lambda/\dAnt$ where $\dAnt$ is the antenna diameter and $\epsilon \approx 0.45$ and that the antennas are Gaussian distributed with a standard deviation of $\sigmaAnt$, equation~\ref{eq:CovVisSuperSimple} allows us to derive a closed-form prediction of the minimal $\kpar$ in such an interferometer that is not contaminated by foregrounds (Appendix~\ref{app:KParMin}),
%\begin{equation}\label{eq:kParMin}
%\kparmin \approx \frac{\sqrt{2}\epsilon \pi}{\freq_0 Y} \frac{\sigmaAnt}{\dAnt} \log \left( \frac{B X^2Y \freq_0 \varR \lambda_0^4}{2 \sqrt{2} \epsilon P_{21} B_{pp} \nAnt k_B^2} \frac{\dAnt}{\sigmaAnt} \right),
%\end{equation}
\begin{align}\label{eq:kParMin}
\kparmin \approx& 1.24\,h\text{Mpc}^{-1} \sqrt{\frac{1+z}{10}} \left(\frac{\sigmaAnt}{1\text{km}}\right) \left(\frac{\dAnt}{10\text{m}}\right)^{-1} \times \nonumber \\
\Big[ 1& + 0.35\log\left(\frac{1+z}{10}\right) -0.04\log\left(\frac{\Omega_m}{0.27}\right) \nonumber \\
&+0.1\log\left(\frac{\sMin}{10\text{mJy}}\right) - 0.08\log\left(\frac{P_{21}}{10^4\text{mK}^2h^{-3}\text{Mpc}^3}\right) \nonumber \\
&-0.08\log\left(\frac{\nAnt}{100}\right) - 0.08\log\left(\frac{\sigmaAnt}{1\text{km}}\right) \nonumber \\
& + \log\left(\frac{\dAnt}{10\text{m}}\right) \Big]
\end{align}
where $P_{21}$ is the amplitude of the 21\,cm power spectrum. This formula can be used to get a quick order-of-magnitude sense as to whether a mode will be accessible to an instrument however it is very optimistic in that it assumes a Gaussian primary beam. While it also strictly assumes that the antennas are distributed as a Gaussian, we have found that it holds to $10\%$ accuracy for non-Gaussian arrays (such as LOFAR and the MWA) as well. 

\begin{figure*}
\includegraphics[width=\textwidth]{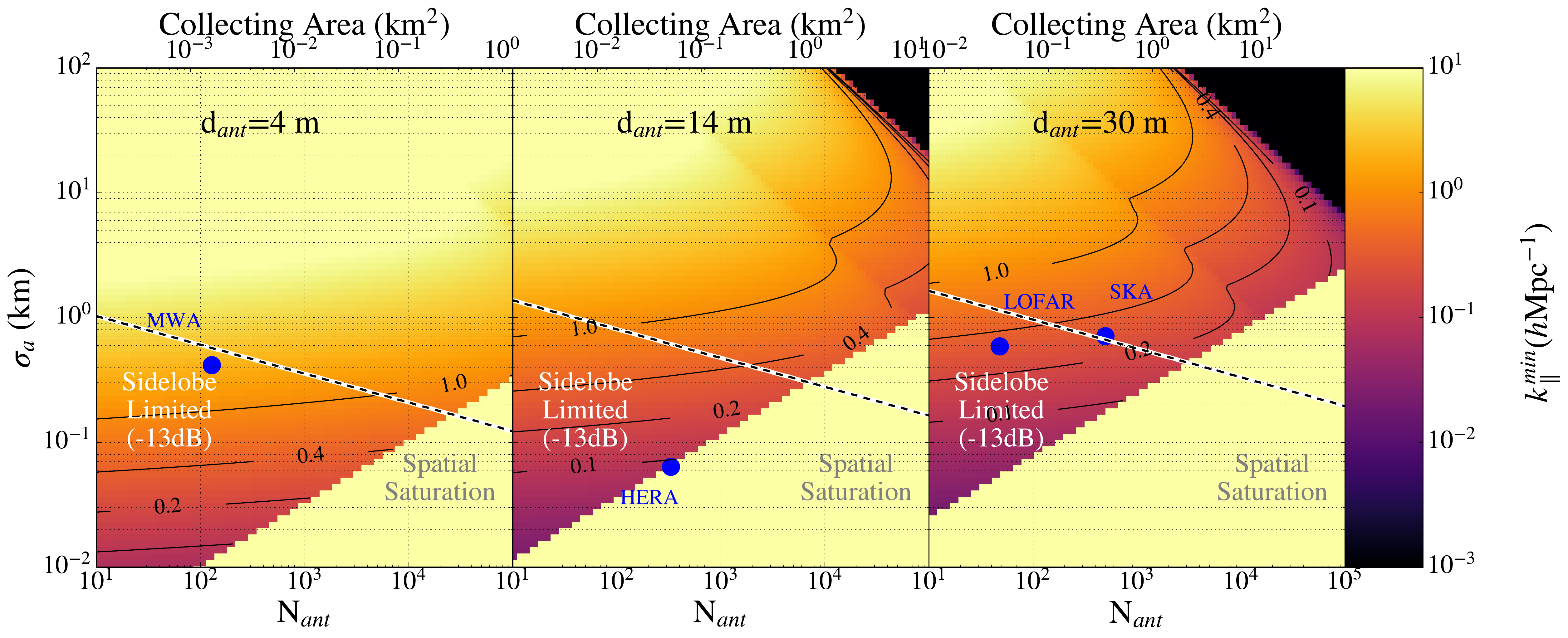}
\caption{Using equation~\ref{eq:kParMin}, we show the smallest $\kparmin$ that is not dominated by modeling noise errors for arrays with Gaussian primary beams, a random circular Gaussian distribution of $\nAnt$ antennas with standard deviation $\sigmaAnt$, each with diameter $\dAnt$.  The area below the white dashed line is where side-lobes render the assumption of Gaussian beams inaccurate. Beige regions on the lower-right hand corner of each plot denote unphysically high packing densities. We see that for all existing instrument designs, calibration noise extends to large $\kpar$ values that will reduce their sensitivity to the 21\,cm signal. HERA will benefit greatly from the fact that it can be calibrated redundantly with minimal reliance on a sky model.}
\label{fig:kParMinConfusion}
\end{figure*}

From equation~\ref{eq:kParMin}, we see that the extent of modeling noise contamination depends primarily on $\sigmaAnt/\dAnt$ while other quantities, such as $\nAnt$ and $\sMin$, are contained within a logarithm have a much weaker impact on $\kparmin$. This proportionality makes sense intuitively since larger apertures have smaller primary beams, suppressing emission at large zenith angles and larger delay. In close packed arrays, the $\sigmaAnt/\dAnt$ proportionality can be saturated so that the $\sigma_r^2$ inside of the logarithm will matter. While this equation ignores the existence of side-lobes, it gives us an order of magnitude estimate of how modeling noise scales with array properties. In Fig.~\ref{fig:kParMinConfusion}, the $\kparmin$ values predicted from the naturally-weighted confusion limits of various planned arrays exceeds $\kparmin\gtrsim 0.2 h$Mpc$^{-1}$, including for the SKA-1 core. Since interferometers such as the SKA and LOFAR focus most of their sensitivity at small $\kpar$ values, their ability to detect the 21\,cm signal will be heavily impacted by foreground modeling errors \citep{Pober:2014}.  

We can get a conservative sense for how side-lobes extend $\kparmin$ beyond the values predicted in equation~\ref{eq:kParMin} by setting the amplitude of the modeling noise at zero-delay, multiplied by square of the side-lobe amplitudes (for an Airy beam, equal to -13dB) equal to the 21\,cm signal (see Appendix~\ref{app:KParMin}). We denote the region of instrumental parameters space that is affected by side-lobes in Fig~\ref{fig:kParMinConfusion} with a grey overlay. Since all planned instruments fall within this region, the $\kparmin$ predictions in this figure are actually optimistic. For these arrays, a more detailed calculation of equation~\ref{eq:CovVisSimplified} with realistic side-lobes is necessary. We found in \S~\ref{ssec:NoiseResultsExisting}, with more realistic side-lobes considered, that the $\kparmin$ obtained is indeed significantly larger than predicted by equation~\ref{eq:kParMin}.

\section{Eliminating Modeling Noise with Baseline Weighting}\label{sec:Mitigation}
While optimistic scenarios in foreground characterization may be precise enough to suppress calibration modeling noise below the 21\,cm signal, elimination of this contamination will also require beam characterization that is beyond the current state of the art. Enabling a power spectrum detection in existing sky-based calibrated experiments calls for an alternate strategy. Redundant calibration is one existing, and so-far successful alternative though it can only be applied to regularly spaced arrays. Though redundant calibration does not rely on a detailed sky model, it is possible that antenna-to-antenna beam variations and position errors can violate the assumption of redundancy and introduce chromatic artifacts that are similar to the ones we have found for sky-based calibration, a potential shortcoming that is being investigated. One approach is to ensure that the instrument contains no structure in region of $k$-space relevant for 21\,cm studies, allowing for smooth fits that do not contaminate the EoR window \citep{Barry:2016}. This is one of the approaches being adopted by HERA \citep{Neben:2016,EwallWice:2016,Thyagarajan:2016,Patra:2016} and an upgrade to the MWA. In this section, we explore an alternative strategy that can be used even when the bandpass is not already intrinsically smooth. By exponentially suppressing long baselines, sky-based calibration is able to remove fine-frequency structure while avoiding contamination within the EoR window. 

  Supra-horizon contamination from calibration noise arises from the inclusion of longer baselines in calibrating gain solutions that are applied to short baselines, leaking power from large too small $\kperp$. One way of mitigating this source of contamination is to weight the visibilities contributing to each gain solution in a way that dramatically up-weights short baselines over long ones. This can be accomplished by choosing an appropriate $\bW$ matrix in equations~\ref{eq:EtaEstimate} and \ref{eq:PhiEstimate}. In \S~\ref{ssec:GaussianWeighting} we explore the efficacy of using a specific form of baseline weighting to eliminate modeling noise. The use of non-unity weights will result in an increase in thermal noise which we discuss in \S~\ref{ssec:GaussianWeightingSensitivity}.
  
  \subsection{Gaussian Weighting for Sky-Based Calibration}\label{ssec:GaussianWeighting}  
  We explore the performance of a $\bW$ matrix that downweights long baselines with the functional form
\begin{equation}\label{eq:GaussianWeights}
\mathsf{W}_{\alpha \beta} = \begin{cases} \exp \left( -\frac{b_\alpha^2}{2 \sigma_w^2} \right) & \alpha = \beta \\
0 & \alpha \ne \beta. \end{cases}
\end{equation}
This function can result in weighs that vary over a range beyond what is allowed for by numerical precision. We note that generally, the off-diagonal elements of the weighting matrix can mix different baselines (which might be desired if we wished to suppress or emphasize features that co-vary between baselines). Our choice of a diagonal matrix corresponds to simply multiplying each visibility by a different weighting factor with which they will contribute to the sum of squares that is being minimized in determining $\bdwEta$ and $\bdwPhi$.  In order to avoid poorly conditioned matrices, a regularization term is also added equal to the identity multiplied by $10^{-6}$, which is large enough to avoid numerical precision errors, but also small enough such that the weights on long baselines are negligible compared to the short ones (and below the dynamic range between foregrounds and signal). With this weighting, core-antennas participating in many short baselines will have their gain solutions dominated by relatively achromatic core visibilities. Meanwhile, outrigger antennas that participate in only long-baselines will derive their solutions from many baselines with similarly small weights. In both cases, a normalization step of $(\bA \bW \bAt)^{-1}$ corrects for the fact that these weights do not sum to unity. Thus, long and short baselines are both effectively calibrated using the Gaussian weighting scheme while the leakage of chromatic errors on long baselines into gain solutions being applied to short baselines is stymied.

\begin{figure*}
\includegraphics[width=\textwidth]{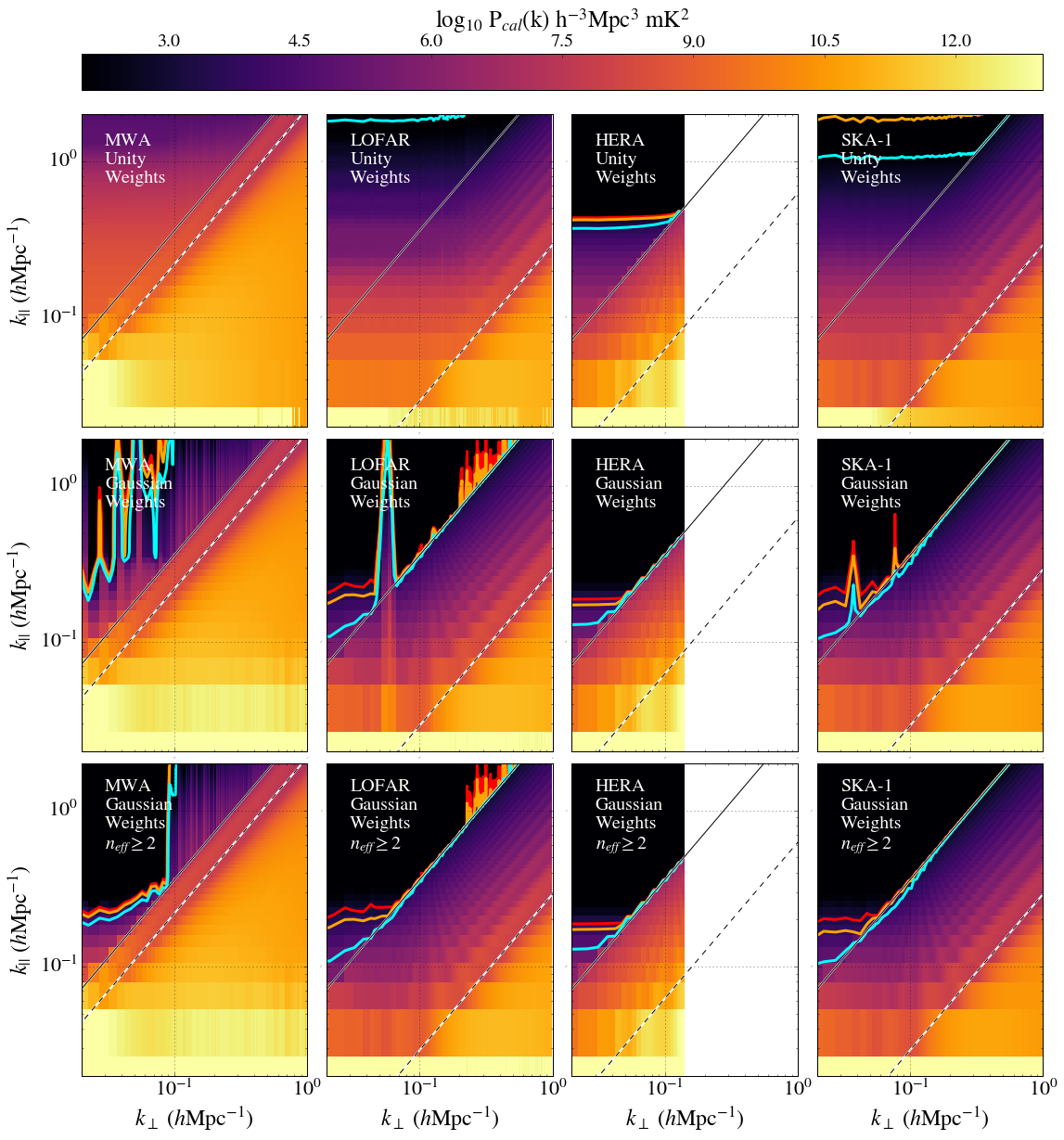}
\caption{Top: Residual power spectra with each visibility weighted equally in determining the calibration solutions ($\bW$ set to the identity matrix). Middle row: the same but now weighting visibilities with a Gaussian function of baseline length (equation~\ref{eq:GaussianWeights}). Much of the EoR window is cleared of contamination from calibration residuals. However pronounced stripes of contamination still exist, especially for LOFAR and the MWA. These stripes arise from short baselines formed from antennas involved in no other short baselines. In order to solve for both antenna gains, they must use information from long baselines, resulting in significant chromaticity on the few short baselines to which the problematic antenna gains are applied.  Bottom: flagging visibilities after calibration until all gains participate in $\Neff \ge 2$ baselines, we find the EoR window free of these stripes. To reiterate, solid lines demarcate regions where the fiducial EoR signal is 1, 5, or 10 times the power of the calibration modeling error. The dashed diagonal line indicates the location of the wedge associated with the first null of the primary beam; the solid line indicates the horizon wedge.}
\label{fig:calWedgesWeighted}
\end{figure*}

We calculate $\mathsf{P}_{\alpha \alpha}$ given by equation~\ref{eq:CovVisSimplified} for the arrays considered in this paper with different values of $\sigma_w$. For LOFAR and the SKA we use $\sigma_w = 100$\,m. For the MWA and HERA, whose cores are especially compact and have larger fields of view than LOFAR and the SKA, we apply more agressive weighting with $\sigma_w=50$\,m. We compare the cylindrically binned and averaged results in the middle row of Fig.~\ref{fig:calWedgesWeighted} with cylindrical power spectra with $\bW$ equal to the identity (top row) and find that most of the EoR window is now free of foreground contamination with the power spectrum accessible at $\kpar \gtrsim 0.1$\,$h$Mpc$^{-1}$ for most arrays.

However, stripes of foreground contamination still extend into the EoR window at distinct $\kperp$ values in the MWA, LOFAR, and to a lesser degree for the SKA. Isolating these baselines in the $uv$ plane, we find that this contamination arises from antennas that are associated with less than two baselines that receive significant weighting.  Define $\Neff(i)$ for the $i^{th}$ antenna to be equal to the sum of the weights of all visibilities that include this antenna divided by their maximum value. As far as calibration is concerned, $\Neff(i)$ describes the effective number of baselines that an antenna participates in. If the number of effective baselines that are used to derive gain solutions is too small, the system is under-constrained and a degeneracy exists between possible solutions for the antenna gains. Two antennas have two gains to solve for, but only one visibility between them. The estimator is forced to break these degeneracies by up-weighting the contribution from the long baselines. An example where $\Neff$ is smaller than two, in our Gaussian weighting scheme, would be for an antenna that is extremely far away from all but one other antenna. Only a single baseline associated with this antenna  has significant weight while the rest are downweighted to zero. 

\begin{figure*}
\includegraphics[width=\textwidth]{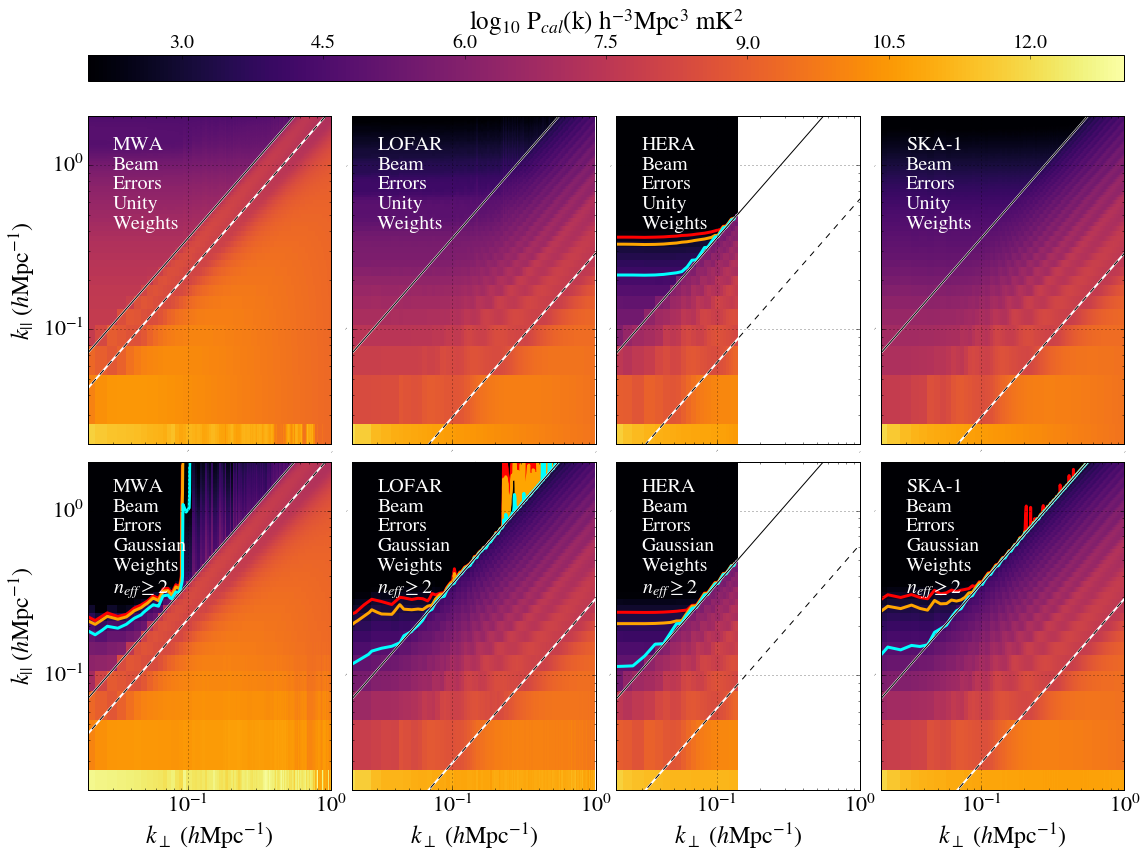}
\caption{Top: Cylinderically binned power spectra of calibration errors due to beam modeling errors at the level of $1\%$ in the main-lobe and $10\%$ in the side-lobes (equation~\ref{eq:BeamErrorModel}). Each visibility has been weighted equally in determining the calibration solutions. Bottom row: The level of cylindrically binned power spectrum residuals from the primary-beam modeling errors in the top row but now with calibration solutions derived from visibilities that are weighted with equation~\ref{eq:GaussianWeights}. Short baselines contributing to antennas with $\Neff \le 2$ have also been flagged from the calibration fit. Weighting with a Gaussian is capable for removing calibration modeling errors due to beam mismodeling at the level that we see in today's experiments \citep{Neben:2015}.}
\label{fig:calWedgesWeightedBeams}
\end{figure*}

We identify these problematic baselines by calculating $\Neff(i)$ for each antenn. We then flag and exclude from the fit the highest weighted visibilities on all antennas with $\Neff\le2$ until all $\Neff$ are greater than $2$. Flagging these visibilities leads to a loss in $\approx 6\%$ of visibilities for LOFAR, $1.3\%$ for the MWA, $0.1\%$ for the SKA, and no visibilities for HERA. The high $\Neff$ for HERA antennas is something we would expect given its compact configuration (every antenna has many short baselines associated with it). Similarly, the SKA core we model is a compact Gaussian with few isolated antennas. LOFAR, on the other hand, has antennas that are arranged in pairs that are separated by short distances so that all of the isolated outriggers have a single short baseline associated with them (which results in the vertical stripe at $\kperp\approx 0.4\,h$Mpc$^{-1}$ in the second row of Fig.~\ref{fig:calWedgesWeighted}). The MWA lacks these pairs, and as a result has fewer low $\Neff$ antennas which tend to lie in the transition between its compact core and extended outriggers. We show cylindrically binned power spectra formed from the delay transform residuals of unflagged visibilities in the bottom row of Fig.~\ref{fig:calWedgesWeighted}, finding that upon flagging this small population of problematic baselines, the EoR window is almost entirely clear above $0.1$\,$h$Mpc$^{-1}$ for all arrays studied. We also show the delay-transformed power spectrum estimates of visibilities contaminated by primary beam modeling errors of $1$\% at zenith and $10$\% in the side-lobes with and without Gaussian visibility weighting applied in the calibration solutions (Fig.~\ref{fig:calWedgesWeightedBeams}). With Gaussian weighting, we are also able to mitigate contamination with the current level of primary beam modeling errors.

\subsection{The Impact of Inverse Baseline Weighting on Power Spectrum Sensitivity}\label{ssec:GaussianWeightingSensitivity}
For an interferometer with identical antenna elements, the thermal noise level on every baseline is the same and $\bN$ is proportional to the identity matrix. For the point source approximation of the modeled foregrounds, the optimal weighting minimizing the errors due to thermal noise in each gain solution is therefore also the identity matrix. Because of its departure from identity weights, the Gaussian weighting that we proposed in the previous section has the effect of increasing thermal noise uncertainties in both the gains and the final power spectrum estimate. In order to see how Gaussian weighting increases the variance due to thermal noise in the gain solutions, one can consider the fact that the variance of the gain solutions goes as $\nAnt^{-1}$ (equation~\ref{eq:CovVisSuperSimple}). For a particular antenna gain, Gaussian weighting reduces the effective number of visibilities whose noises are averaged over in each gain solution so that the variance of the antenna gain is now $\sim \Neff^{-1}$ rather than $\nAnt^{-1}$. In the weighting schemes employed in \S~\ref{ssec:GaussianWeighting}, $\Neff$ goes down by a factor of order $1-10$, remaining between $10-100$ for LOFAR and the MWA.% The errors on each delay transformed visibility may be determined. 

Assuming Gaussian errors, the covariance between the square of two delay-transformed visibilities is given by 
\begin{align}
\sigma^2_{\alpha\beta} &= \left \langle | \ftV_\alpha(\tau) |^2 |\ftV_\beta(\tau)|^2 \right \rangle - \left \langle | \ftV_\alpha(\tau) |^2 \right \rangle \left \langle | \ftV_\beta(\tau) |^2 \right \rangle \nonumber \\
& = \left( \mathsf{P}^N_{\alpha \beta} + \mathsf{P}^R_{\alpha \beta} +\mathsf{P}^S_{\alpha \beta} \right)^2
\end{align}
where $\mathsf{P}^N_{\alpha \beta} \equiv \left \langle \ftV_\alpha^N(\tau) \ftV_\beta^{N*}(\tau) \right \rangle$ is the covariance matrix of the thermal noise component of delay-transformed visibilities and $\mathsf{P}^{R(S)}_{\alpha \beta}$ are the covariances of the delay-transformed residual foreground (signal) visibilities. While the residual foreground component can contribute significantly, it is only of concern in the regions of $k$-space where the amplitude of the foreground modeling noise is comparable to or greater than the level of the 21\,cm signal. Since we are interested in how the thermal noise increases in the region of k-space where we have reduced foreground bias to well below the signal level, we will focus our attention on the thermal noise component and ignore the sample variance from modeling noise and signal for the remainder of this discussion.

We may compute $\mathsf{P}^N_{\alpha \beta}$ using equation~\ref{eq:CovVisSimplified} with $\cC \to \cN$. Typically, thermal noise is uncorrelated between baselines so $\mathsf{\widetilde{N}}^{\alpha \beta}$ is diagonal. In the absence of calibration errors, the covariance between the squares of different delay-transformed visibility products arising from thermal noise would therefore also be zero. The presence of calibration errors introduces additional components to the thermal noise in all but the last term of equation~\ref{eq:CovVisSimplified} that are correlated from baseline to baseline. For identity weights and a diagonal noise-covariance, the off-diagonal terms in $\mathsf{P}_{\alpha \beta}$ go roughly as $\sim \nAnt^{-2} \sim \nVis^{-1}$ compared to the diagonal terms (which have the order unity contribution that does not arise from calibration). Thus, for $\alpha \ne \beta$, $\sigma^2_{\alpha \beta} \sim \nVis^{-2} \sigma^2_{\alpha \alpha}$ and has, so far, been ignored in other sensitivity calculations \newpage (e.g. \citealt{McQuinn:2006}, \citealt{Parsons:2012b},  \citealt{Beardsley:2013}, and \citealt{Pober:2014}). 

In order to obtain enough sensitivity for a detection, interferometry experiments are expected to perform spherical binning and averaging in $k$-space to obtain power-spectrum estimates, $\widehat{p}_A$ whose covariance we denote as $\Sigma_{AB}$ (denoting band-powers with upper-case latin subscripts). The variance of a binned and averaged power spectrum estimate with identity weights is given by 
\begin{align}
\Sigma_{AA}  &= N_A^{-2}\left(\sum_{\alpha \in A} \sigma_{\alpha \alpha}^2 + \sum_{\alpha \in A} \left[\sum_{\beta \in A; \alpha \ne \beta} \sigma_{\alpha \beta}^2\right]\right)\nonumber \\
& \sim N_A^{-2} \left( \sum_{\alpha \in A} \sigma_{\alpha \alpha}^2 + \frac{N_A}{\nVis^2} \sum_{\alpha \in A} \sigma_{\alpha \alpha}^2 \right).
\end{align}
Thus, the contribution to $\Sigma_{AA}$ from off-diagonal elements of the noise-covariance is sub-dominant to the  contribution from diagonal elements as $\sim N_A/\nVis^2$ where $N_A$ is the number of visibilities averaged within the $A^{th}$ bin. 
 
Non-uniform weighting in calibration decreases the effective number of visibilities in calibration, increasing the off-diagonal terms in equation~\ref{eq:CovVisSimplified}. This in turn leads to an increase in the overall error bar on each spherically binned and averaged power spectrum estimate. We compute the degree to which Gaussian weighting degrades sensitivity to the spherically binned power spectrum by comparing $\Sigma_{AA}$ for both uniform and Gaussian weighting within a single LST. While calibration correlates the noise on different squared visibilities in the same power spectrum bin, we can minimize the extra error by inverse-covariance weighting them before averaging. 

\begin{figure}
\includegraphics[width=.47\textwidth]{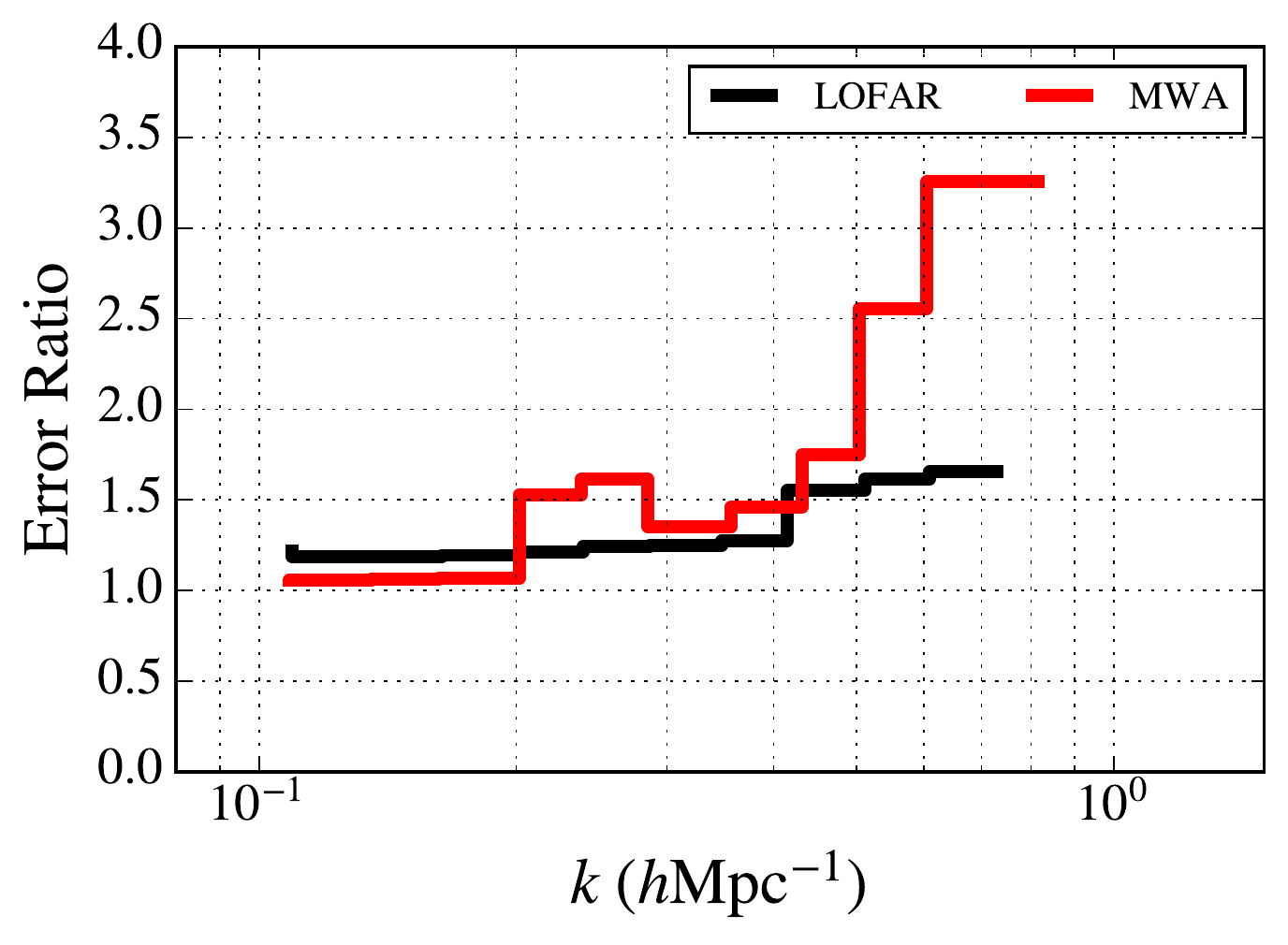} \vspace{-5pt}
\caption{The ratio between thermal noise errors on a spherically averaged power spectrum estimate with Gaussian weighting and uniform weighting of visibilities in calibration. We show this ratio for LOFAR where $\sigma_w=100$\,m and the MWA with $\sigma_w=50$\,m. In both cases, the reduction in sensitivity to the power spectrum is by a factor less than two for small $k$ where the interferometers have maximum sensitivity. Under the Gaussian weighting scheme, antennas with fewer short baselines have increased thermal noise in their gains. Increasingly large k-bins include larger numbers of visibilities formed from antennas with fewer short baselines (small $\Neff$) which have large increases in their thermal noise, leading to a trend of increasing sensitivity loss with increasing $k$. Since the MWA has a narrower weighting function, with $\sigma_w=50$\,m, this increase occurs faster than for LOFAR.}\label{fig:WeightedNoiseRatio}
\end{figure}

We perform this averaging and report how the Gaussian down-weighting of long baselines affects the thermal noise on the final power spectrum estimate in Fig.~\ref{fig:WeightedNoiseRatio} . Because the covariance matrices for HERA and the SKA are very-large and would require significant computation to invert, we only perform this calculation for LOFAR and the MWA. We also assume that each power spectrum estimate only incorporates visibilities outside of the wedge. The proportion of long baselines which tend to be formed from antennas with smaller $\Neff$ increases with each $k$-bin. Hence, the decrease in sensitivity increases with $k$. Since the MWA weighting function is more compact, with $\sigma_w=50$\,m, the increase in the error ratio goes faster than for LOFAR which has a wider weighting function with $\sigma_w=100$\,m. Within the region that instruments are expected to be sensitive to the 21\,cm signal, the error bars only go up by less than two. Gaussian weighting increases the thermal noise in the power spectrum measurement, but only by a level similar to intrinsic thermal noise that would be present even if calibration were perfect. Gaussian weighting can therefore allow us to circumvent the problem of foreground modeling noise in calibration while only sacrificing a small amount of sensitivity to the 21\,cm power spectrum.

While baseline-dependent weighting is able to clear the EoR window, it does not necessarily allow any instrument to work within the wedge. Figs.~\ref{fig:CompareModelNoise} and \ref{fig:BeamErrors} show that this would still require superb foreground models accurate to the $0.1$\,mJy level and modeling of the primary beam to the $10^{-3}$ level in the main lobe and $10^{-2}$ level in the side-lobes. Until these milestones are achieved, extended arrays will suffer a disproportionate reduction in delay power-spectrum sensitivity relative to compact arrays like HERA \citep{Pober:2014}.

\section{Conclusions}\label{sec:Conclusion}

In this work, we derived expressions for the amplitude of the power spectrum bias arising from the imprint of foreground modeling errors on calibration. These expressions assumed that calibration errors are small enough such that their solutions are obtained through a linear set of equations, which is the case in the final stages of iterative, sky-based calibration schemes when the errors in the foreground model are small. Using these equations we are able to explain the amplitude of the biases that have been simulated for the special cases of the MWA (B16) and LOFAR \citep{Patil:2016} and to predict the amplitude of modeling noise in the power spectrum for the SKA-1 and HERA (which does not actually rely on this approach). We performed this analysis in a variety of foreground and beam modeling scenarios. We also use our formalism to determine the dependence of modeling noise on the parameters of the array and the accuracy of the calibration catalog. These results do not apply to the redundant calibration strategies used by HERA and PAPER, although errors introduced by deviations from redundancy still have the potential to contaminate the window in a similar way. Our analysis also reveals that noise bias exists in current power spectrum estimates where separate calibration solutions are not obtained for interleaved data sets. Whether this bias limits 21\,cm experiments requires further analysis but it can easily be avoided by obtaining independent calibration solutions for cross multiplied data. 

This paper aimed to illuminate the source of calibration errors within the EoR window. In order to make our analysis analytically tractable, we employed a number of assumptions. These include assuming that the array is minimally redundant so that we can ignore off diagonal  elements of the visibility covariance matrix, and that the sources themselves are flat-spectrum. A more significant assumption that will not hold in many observing scenarios is that we ignored the chromaticity of modeled foregrounds, which holds approximately when the modeled fluxes are dominated by a source at the phase center that exceeds the flux of the next brightest source by a factor of a few. We also assumed that our instruments had Airy-beams, that sources could be characterized down to a fixed flux-level across the entire sky, ignored ionospheric effects and polarization \citep{Sault:1996,Jelic:2010,Moore:2013,Asad:2015,Kohn:2016,Moore:2017} which is especially severe on the large spatial scales \citep{Lenc:2016} that we suggest should be relied upon in calibration strategies. Hence, specific quantitative predictions in this paper should be regarded as accurate to within an order of magnitude and on the optimistic side. In validating the design of a future instruments, full end-to-end simulations should be employed, though this is left to future work.

Our calculations indicate that for current catalog limits presented in \citet{Caroll:2016,HurleyWalker:2016}, and \citet{Williams:2016} both the MWA and LOFAR will observe an EoR window that is heavily contaminated by chromatic calibration errors due to unmodeled sources. Since the chromaticity of these errors increases with the length of baselines involved in calibration, removing inner baselines from calibration, as is required to avoid signal loss with direction dependent calibration \citep{Patil:2016} will only exacerbate these chromatic errors and is probably the source of the systematics floor observed by LOFAR in \citet{Patil:2017} (these authors note that calibration errors as a likely culprit but not that the use of long baselines is exacerbating the problem).  Our analytic treatment suggests that instead, sky-based experiments should use their short baselines to calibrate power-spectrum data which may preclude the direction-dependent approach to avoid signal loss and will likely require more accurate models of diffuse emission. LOFAR may also be able reduce the amplitude of calibration errors below the power spectrum, at large spatial scales, by averaging over multiple fields of view (if its gains are temporally stable) and/or by building a source catalog complete down to $\approx 100\,\mu$Jy across the entire sky. Even if such a catalog is constructed, beam modeling precision will also need to be improved by an order of magnitude over what has been achieved in the literature. The large field of view on the MWA decreases the number of fields that can be averaged over and increases the $\kpar$ values contaminated by modeling errors, making the path to removing this noise with extant methods considerably more difficult than for LOFAR.

Our analysis motivates a potential solution to the problem of modeling noise in sky-based calibration. Since contamination within the EoR window arises from the coupling of long baseline errors into the calibration solutions on short ones, our proposed strategy is to down-weight the contribution of long baselines to the gain solutions that are applied to short baselines. The linear least-squares estimator formalism employed in this paper provides a natural framework for incorporating such weights. Experimenting with a Gaussian weighting scheme, we find that down-weighting long baselines should allow for both existing and future arrays to correct fine-frequency bandpass structures without introducing chromatic sky-modeling errors. While such weighting will increase the level of thermal noise present in calibration solutions, we find that this noise increase will only result in power spectrum error bars that are $\approx 1-1.5$ times larger than the case where all visibilities are weighted identically. This method prevents calibration errors from limiting the foreground avoidance approach, which seeks to detect the 21\,cm signal within the EoR window and thus requires the calibrated instrumental response to be spectrally smooth. This method is not sufficient to enable {\it foregrounds subtraction}; accessing the signal inside of the wedge. Working within the delay-wedge will require significant improvements in foreground and primary beam modeling.

\section*{Acknowledgements} 
We would like to thank Danny Jacobs, Nichole Barry, Miguel Morales, Jonathan Pober, Bryna Hazelton, Cathryn Trott, and Aaron Parsons for helpful discussions.
A.E.W. acknowledges support from an NSF Graduate Research Fellowship under Grant No. 1122374. J.S.D. acknowledges support from a Berkeley Center for Cosmological Physics Fellowship. A.L. acknowledges support for this work by NASA through Hubble Fellowship grant \#HST-HF2-51363.001-A awarded by the Space Telescope Science Institute, which is operated by the Association of Universities for Research in Astronomy, Inc., for NASA, under contract NAS5-26555.
A portion of this work was performed at the Aspen Center for Physics, which is supported by National Science Foundation grant PHY-1066293.
\bibliographystyle{mnras}
\bibliography{calNoise}

\appendix

\section{The Impact of Redundancy}\label{app:Redundancy}
Throughout this paper, we ignored the  impact of redundancy between visibilities, letting $\mathsf{R}_{\alpha \beta}$ be diagonal when calculating modeling noise. However, redundancy is significant in highly compact arrays, such as HERA. Here we argue that the impact of redundancy on the modeling noise levels, calculated in this work, is to multiply the overall noise level by a factor of order unity which only has a small effect on the extent of contaminated modes in $k$-space. We also verify this argument with a numerical calculation.

When $\mathsf{C}$ is diagonal, the sum in equation~\ref{eq:VisibilityCovApprox} is only over terms with $\gamma = \delta$. The existence of redundant baselines introduces non-negligible off-diagonal terms in the visibility covariance matrix $\bR$. For each $\gamma=\delta$ term in the non-redundant sum, we can consider the additional summands, with $\gamma \ne \delta$ that are introduced for each $ii/jj$ term and $ij/ji$ term. We start with $ii/jj$.

For a fixed baseline $\gamma$ that involves antenna $i$, there will be at most $\sim \nAnt$ additional baselines that are redundant with $\gamma$ and do not involve gain $i$. From equations~\ref{eq:etaWeights} and \ref{eq:phiWeights}, the weighting of the covariance between two different baselines in which only one involves antenna $i$ goes as $\sim \nAnt^{-3}$. Thus, the presence of redundancy adds no more than $\sim \nAnt$ terms involving antenna $i$ but not antenna $j$ and vice versa, for each $ii$ summand in equation~\ref{eq:VisibilityCovApprox}. Multiplying this overall factor of $\nAnt^{-2}$ by $\nAnt$ to account for the $\nAnt$ different $ii$ sums leads to a contribution to the noise amplitude on the order of $\sim  \nAnt^{-1}$, similar to the level of the noise without redundancy. As a result, redundancy changes the modeling error amplitude by a factor of order unity in the diagonal terms. Next we consider the $ij$ terms in equation~\ref{eq:VisibilityCovApprox}. 

For a given $\gamma$ and $i\ne j$, there will be at most $\sim \nAnt$ redundant baselines that do not involve the $i^{th}$ or $j^{th}$ gains, causing the weighting of each unique variance term to go as $\nAnt^{-3}$ rather than $\nAnt^{-4}$ in the non-redundant case. Since there are $\sim \nAnt^2$ unique baselines that do not involve $i$ or $j$, the overall sum of these terms goes as $\nAnt^{-1}$. As a result, the $ij$ terms in equation~\ref{eq:VisibilityCovApprox} will have a similar magnitude as the $ii/jj$ terms but the overall impact on the amplitude of the modeling noise described in equation~\ref{eq:VisibilityCovApprox} still changes the amplitude by a factor of order unity.

\begin{figure}
\includegraphics[width=.5\textwidth]{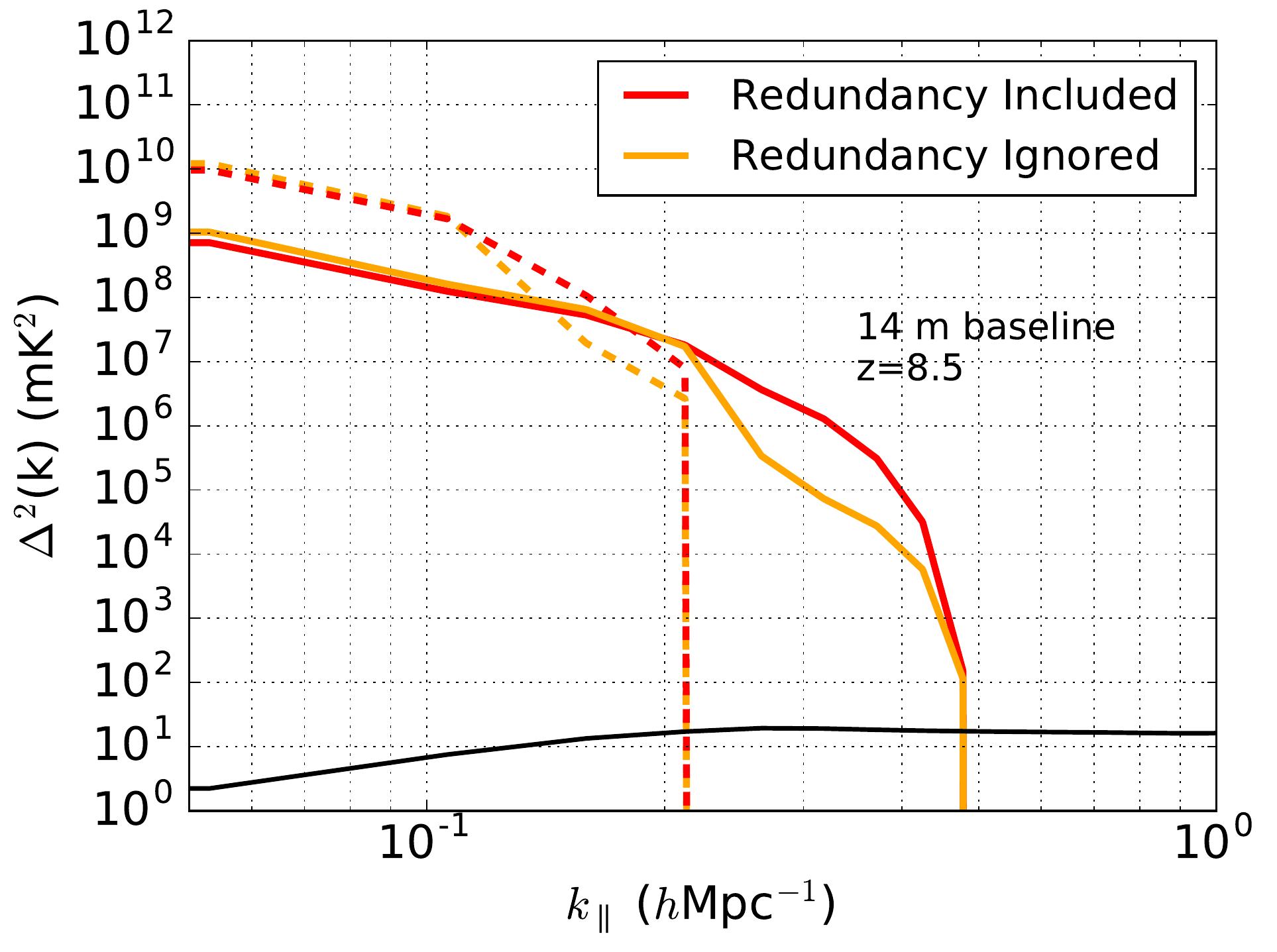}
\caption{We compare the amplitude of modeling noise on a short baseline when $\bR$ is assumed to be diagonal (orange line) and the off-diagonal terms of $\bR$ are explicitly included for a 91-element (dashed lines) and 331-element (solid lines) hexagonally packed array of 14\,m apertures. We find that even in a maximally redundant array, the effect of redundancy is to change the overall amplitude of the modeling noise by a factor of a few. This only has a small impact $(\lesssim 10\%)$ on $\kparmin$, the smallest $\kpar$ where the 21\,cm signal (black line) dominates over the modeling noise, as computed from equation~\ref{eq:kParMin} which ignores the effect of redundancy.}
\label{fig:RedundancyCompare}
\end{figure}

We confirm these arguments with a numerical comparison between the amplitude of the modeling noise with and without redundancy taken into account for two redundant arrays of 91 and 331 hexagonally packed $14$\,m apertures and $\sMin$ equal to the naturally-weighted confusion limit. We compute the off-diagonal elements of $\bR$ by numerically computing the beam integral in equation~\ref{eq:VisCovariancePoints} for all $\cR_{\alpha\beta}$ with Airy beams and perform the full matrix inversions prescribed in equation~\ref{eq:EtaEstimate} and \ref{eq:PhiEstimate}. We compare our results to the same calculation where all off-diagonal elements of $\bR$ are set to zero (Fig.~\ref{fig:RedundancyCompare}) and find that the difference in amplitude is essentially a factor of order unity, leading to a negligible increase in the effective $\kparmin$. This calculation confirms our argument for HERA-scale arrays. 

%Redundancy_Compare_Hex_91.ipynb

\section{The Point Source Approximation For Modeled Foregrounds.}\label{app:ModelApprox}
For analytic tractability, we assumed that the modeled component of foregrounds were well characterized by a flat-spectrum point source at zenith, whose visibilities are achromatic. Throughout the paper, the rest of the unmodeled foregrounds considered in our calculations were not assumed to be a single point source and are characterized by chromatic visibilities (see \S~\ref{sssec:PointSources}-\ref{sssec:Diffuse}). In this appendix we explore the consequences of relaxing this assumption. 

A significant consequence of the foregrounds not being dominated by a single point source at the phase center  is that for a fixed frequency, $y_\alpha$'s amplitude will vary significantly from baseline to baseline, often approaching zero where source fringes destructively interfere. As a result, $\cov \left[\frac{c_\alpha}{y_\alpha}, \frac{c_\alpha^*}{y_\alpha^*} \right]$ can vary rapidly in frequency where $y_\alpha$ approaches zero. Thus, any weighting scheme that does not take these nulls into account will experience calibration error chromaticity in large excess of what we have found so far. 

Instead, it is typical for calibration solutions to be obtained for each frequency through inverse covariance weighting. Since the thermal-noise covariance matrix is usually proportional to the identity,  per-frequency inverse covariance weights are proportional to $|y_\alpha|^2$. Under this scheme, we may employ a weights matrix that is frequency dependent. 
\begin{equation}
\mathsf{W}_{\alpha \alpha} \to \mathsf{W}'_{\alpha \alpha}= \mathsf{W}_{\alpha \alpha}|y_\alpha(\freq)|^2
\end{equation}
which leads $\Lm$ and $\Pm$ to be frequency dependent as well and we can no longer separate them from $\ftC$ in the delay-transform. The delay-transform visibility in equation~\ref{eq:tMultiply} becomes
\begin{equation}
\ftV_\alpha \approx \int d\freq e^{2 \pi i \freq \tau} \left[ y_\alpha  \left( \feta_i' + \feta_j' + i \fphi_i' - i \fphi_j' \right) + c_\alpha \right]
\end{equation} 
where every term, including $y_\alpha$ is a function of frequency. The expectation value for the Delay-transformed product of $\ftV_\alpha$ with its complex conjugate to second order in $\bc/\by$ (equation~\ref{eq:CovVis}) is now,
\begin{align}\label{eq:CovChromatic}
\mathsf{P}_{\alpha \beta}  = \int d\freq d \freq' e^{2 \pi i \tau(\freq-\freq')} [ y_\alpha y_\beta^*\langle \eta_i'\eta_\ell^{\prime*} \rangle &+ y_\alpha y_\beta^* \langle \eta_j' \eta_m^{\prime *} \rangle \nonumber \\
 + y_\alpha y_\beta^* \langle \eta_j'\eta^{\prime *}_\ell \rangle & + y_\alpha y_\beta^* \langle \eta_i'\eta_m^{\prime*}\rangle \nonumber \\
+ y_\alpha y_\beta^* \langle \phi_i' \phi_\ell^{\prime*} \rangle & + y_\alpha y_\beta^* \langle \phi_j' \phi_m^{\prime *} \rangle \nonumber \\
-i y_\alpha y_\beta^* \langle \phi_i' \phi_m^{\prime *} \rangle & - i y_\alpha y_\beta^* \langle \phi_j' \phi_\ell^{\prime *} \rangle \nonumber \\
+ y_\alpha \langle \eta_i' c_\beta^{\prime *} \rangle & + y_\alpha \langle \eta_j' c_\beta^* \rangle  \nonumber \\
+ i y_\alpha \langle \phi_i' c_\beta^* \rangle & - i y_\alpha \langle \phi_j' c_\beta^* \rangle  \nonumber \\
+ y_\beta^* \langle \eta_\ell'^* c_\alpha \rangle & + y_\beta^* \langle \eta_m' c_\alpha \rangle \nonumber \\
- i y_\beta^* \langle \phi_\ell^{\prime *} c_\alpha \rangle & + i y_\beta^* \langle \phi_m^{\prime *} c_\alpha \rangle \nonumber \\
& + \langle c_\alpha c_\beta^* \rangle. ]
\end{align}
where every complex conjugated quantity is a function of $\freq'$ and every non-conjugated quantity is a function of $\freq$. Since the weight and design matrices are no longer frequency independent, second order moments cannot be sepearated into frequency dependent and independent components as we did with the point source approximation. In order to compute $\mathsf{P}_{\alpha \beta}$, we must calculate all second order moments with a given source model and design matrix and take the Fourier transforms. 

For realistic $y_\alpha$, we use simulations of point source foregrounds obtained from the PRISim software package \citep{Thyagarajan:2015a} for the MWA-128T array layout with antennas modeled as 4\,m diameter dishes. For each 100\,kHz channel over a 20\,MHz band, we use a weights matrix $W'_{\alpha \alpha}(\freq) = W_{\alpha \alpha} |y_\alpha(\freq)|^2$ and compute the two-dimensional Fourier transform in equation~\ref{eq:CovChromatic} to obtain $\mathsf{P}_{\alpha \alpha}$ for several baselines. We take the MWA to be pointing at a declination equal to its latitude of $-26.701^\circ$ \citep{Tingay:2013a} at LST=0 and 4\,hr.  We run two different simulations, one in which $\mathsf{W}_{\alpha \alpha}$ is set to unity (and the weights matrix $\mathsf{W}_{\alpha \alpha}'=\mathsf{W}_{\alpha \alpha} |y_\alpha|^2$)  and the other where $\mathsf{W}_{\alpha \alpha}$ is given by equation~\ref{eq:GaussianWeights} with $\sigma_w=50$\,m. 

\begin{figure*}
\includegraphics[width=\textwidth]{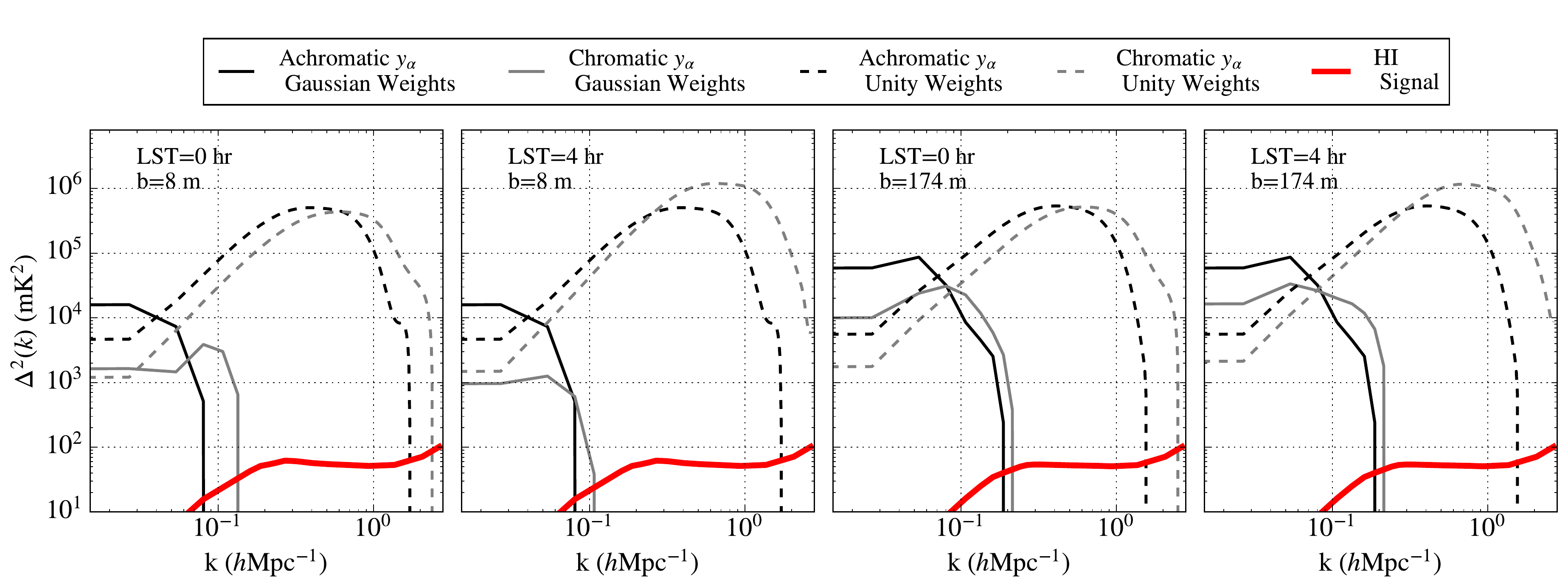}
\caption{Comparisons between calibration modeling noise with realistic  modeled foregrounds (grey lines) and the point source foregrounds used throughout this paper (black lines) with (solid) and without (dashed) Gaussian baseline weighting for two different LSTs and baseline lengths on the MWA from 10\,MHz noise equivalent bandwidth centered at 150\,MHz. The red line denotes the amplitude of the HI power spectrum generated with 21cmFAST. We find that fully modeled foregrounds change the overall amplitude of the of the calibration noise since the amplitude of a particular modeled visibility does not necessarily equal the amplitudes of every other modeled visibility. Chromatic $y_\alpha$s also introduce some additional spectral structure which results in a larger width of calibration errors in $k_\parallel$. The overall impact on the LoS mode where modeling noise bias falls below the 21\,cm signal is only on the order of $10\,\%$ with Gaussian weighting.  }
\label{fig:CompareModelNoise}
\end{figure*}

In Fig.~\ref{fig:CompareModelNoise}, we show the amplitude of calibration modeling noise on a short (8\,m) and  long (174\,m) MWA baseline with and without fully modeled foregrounds for several different LSTs. In all cases, we see that the fully modeled foregrounds extend the width of the foreground noise to larger $\kpar$, something we would expect to occur with the additional spectral structure they introduce. In addition, the amplitude of the modeling noise is modified since the multiplication by the modeled foregrounds on a particular baseline, $y_\alpha y_\beta^*$ (equation~\ref{eq:CovChromatic}) does not necessarily cancel out the modeled foregrounds in the numerator of each summed $\langle c_\gamma c_\delta^* \rangle/(y_\gamma y^*_\delta)$ as they do when $y_\alpha$ is constant. Despite these differences, we find that over the range of LSTs and baselines studied, the overall impact on the minimal LoS wavenumber of modes that can be observed at a particular $uv$ point is only on the order of $\approx10\,\%$. Thus, the approximation of the modeled foregrounds as a point-source at the phase center does not have a significant impact on range of modes that are masked by foreground modeling errors.

\section{Expressions for Second Moments of Delay-Transformed Calibration Errors.}\label{app:Approximate}
In this section, we derive the approximate expressions for the second moments that we use to go from equation~\ref{eq:CovVis} to equation~\ref{eq:CovVisSimplified}. To derive equations~\ref{eq:Simplification1} through \ref{eq:Simplification5}, we first note that
\begin{align}
\cov[\re(\bc),\re(\bc)^\intercal] &\approx \frac{1}{2} \bC. \nonumber \\
\cov[\im(\bc),\im(\bc)^\intercal] &\approx \frac{1}{2} \bC. 
\end{align}
This assertion is true for thermal noise, $\bn$, since both the real and imaginary components of the thermal noise are given by identical, zero-mean normal distributions. We need only show that this assertion holds for the unmodeled foregrounds $\br$. 
We start by writing
\begin{align}
&\cov[\re(\br),\re(\br)^\intercal]_{\alpha \beta} \nonumber \\
& \propto \int d \Omega \cos\left(\frac{2 \pi \nu \bb_\alpha \cdot \bs}{c} \right) \cos \left( \frac{2 \pi \nu' \bb_\beta \cdot \bs }{c} \right) |A(\bs)|^2 \nonumber \\
& = \frac{1}{4} \int d \Omega \left[ e^{2 \pi i (\freq \bb_\alpha +\freq' \bb_\beta)\cdot \bs/c} + e^{-2 \pi i (\freq \bb_\alpha + \freq' \bb_\beta )\cdot \bs/c}  \right] |A(\bs)|^2\nonumber \\
& + \frac{1}{4} \int d \Omega  \left[ e^{2 \pi i (\freq \bb_\alpha - \freq' \bb_\beta )\cdot \bs /c} + e^{-2 \pi i (\freq \bb_\alpha - \freq' \bb_\beta)\cdot \bs /c} \right]|A(\bs)|^2,\label{eq:RealCov}
\end{align}
where we dropped the multiplicative instrument-independent terms in the covariance (equations~\ref{eq:VisCovariancePoints} \& \ref{eq:VisCovarianceDiffuse}) in favor of a proportionality sign. 
All of the terms in equation~\ref{eq:RealCov} will integrate to zero unless $|\bb_\alpha \pm \bb_\beta| \lesssim \dAnt$ (less than one fringe fits within the primary beam main-lobe) which is only true if $\bb_\alpha \approx \pm \bb_\beta$ where the negative case causes exponential terms in the  first line of equation~\ref{eq:RealCov} to be non-zero and the positive case causes the second line to be non-zero. We may choose baseline indexing such that we never have $\bb_\alpha \approx -\bb_\beta$, by having antenna numbers increase with increasing E-W and than N-S position.  With this indexing,
\begin{align}
&\cov[\re(r_\alpha),\re(r_\beta)] \nonumber \\
& \propto \frac{1}{4} \int d \Omega \left[ e^{2 \pi i (\freq \bb_\alpha - \freq' \bb_\beta )\cdot \bs /c}+e^{-2 \pi i (\freq \bb_\alpha - \freq' \bb_\beta)\cdot \bs /c}  \right]|A(\bs)|^2 \nonumber \\
& = \frac{1}{4} [ \cov(r_\alpha,r_\beta^*) + \cov(r_\alpha^*,r_\beta)].
  \end{align}
  For beams that are symmetric around the phase center, $\cov[ r_\alpha,r_\beta^*]$ is real and $\cov[r_\alpha,r_\beta^*] = \cov[r_\alpha^*,r_\beta]$, proving our assertion that 
\begin{equation}\label{eq:CovRealApprox}
    \cov[\re(r_\alpha),\re(r_\beta)] = \frac{1}{2} \cR_{\alpha\beta}.
\end{equation}
A very similar set of steps with identical assumptions yields
  \begin{equation}
  \cov[\im(r_\alpha),\im(r_\beta)] = \frac{1}{2} \cR_{\alpha\beta}.
  \end{equation}

Next, we show that
\begin{equation}\label{eq:CovDominance}
\cC_{\alpha \beta} \gg \langle c_\alpha \rangle \langle c_\beta^* \rangle. 
\end{equation}
This can be seen by writing the product of the averages
\begin{align}
\langle c_\alpha \rangle \langle c_\beta^* \rangle =&  \int d\Omega e^{-2 \pi i \freq \bb_\alpha \cdot \bs/c} A(\bs) \langle I(\bs) \rangle \nonumber \\
 \times &\int d\Omega' e^{2 \pi i \freq' \bb_\beta \cdot \bs /c} A(\bs') \langle I(\bs')\rangle
\end{align}
both integrate to zero when $b_{\alpha/\beta} \gtrsim \dAnt$ and $\langle I(\bs) \rangle$ is smooth as a function of position (which is typically true of foreground residuals and signal). 

We can now show derivations for equations~\ref{eq:Simplification1} through \ref{eq:Simplification5}. 

\subsection{Derivation of Equation~\ref{eq:Simplification1}}
We start on the left hand side with 
\begin{align}
\langle \feta_i'\feta_j^{\prime *} \rangle &= \int d \freq d \freq' e^{-2 \pi i \tau (\freq-\freq')} \left[ [\bcEta]_{ij} + \langle \eta_i \rangle \langle \eta_j^{\prime *} \rangle  \right]\nonumber \\
& \approx   S_0^{-2}\Lm_{i\gamma}\Lm^\intercal_{\delta j} \int d \freq d \freq' e^{-2 \pi i \tau (\freq-\freq')}\cov[\re(c^\gamma),\re(c^\delta)]\nonumber \\
& = \frac{S_0^{-2}}{2} \Lm_{i \gamma}\Lm_{\delta j}^\intercal \int d \freq d \freq' e^{-2 \pi i \tau (\freq-\freq')} \covC^{\gamma \delta}(\freq,\freq') \nonumber \\
&= \frac{S_0^{-2}}{2} \Lm_{i\gamma}\Lm^\intercal_{\delta j} \ftC^{\gamma \delta}
\end{align}
In going from the first to the second line, we threw away the product of the means (equation~\ref{eq:CovDominance}). Going from the second to the third line, we used equation~\ref{eq:CovRealApprox}.

\subsection{Derivation of Equation~\ref{eq:Simplification2}}
Following the same procedure for equation~\ref{eq:Simplification1},
\begin{align}
\langle \phi_i'\fphi_j^{\prime *} \rangle &= \int d \freq d \freq' e^{-2 \pi i \tau (\freq-\freq')} \left[ [\bcPhi]_{ij} + \langle \phi_i \rangle \langle \phi_j^{\prime *} \rangle  \right]\nonumber \\
& \approx   S_0^{-2}\Pm_{i\gamma}\Pm^\intercal_{\delta j} \int d \freq d \freq' e^{-2 \pi i \tau (\freq-\freq')}\cov[\im(c^\gamma),\im(c^\delta)] \nonumber \\
& = \frac{S_0^{-2}}{2} \Pm_{i \gamma}\Pm_{\delta j}^\intercal \int d \freq d \freq' e^{-2 \pi i \tau (\freq-\freq')} \covC^{\gamma \delta}(\freq,\freq') \nonumber \\
&= \frac{S_0^{-2}}{2} \Pm_{i\gamma}\Pm^\intercal_{\delta j} \ftC^{\gamma \delta}.
\end{align}

\subsection{Derivation of Equation~\ref{eq:Simplification3}}
Starting with the left-hand side of equation~\ref{eq:Simplification3},
\begin{align}
\langle \fc_\alpha \feta_i^* \rangle & = \frac{S_0^{-1}}{2}\int d \freq d \freq' e^{-2 \pi i (\freq - \freq')\tau}\Lm_{i \gamma}\langle c_\alpha (c^{\gamma} + c^{\gamma*})\rangle \nonumber \\
& \approx \frac{S_0^{-1}}{2} \Lm_{i \gamma} \int d \freq d \freq' e^{-2 \pi i(\freq-\freq')\tau} \covC_\alpha{}^\gamma(\freq,\freq') \nonumber \\
& = \frac{S_0^{-1}}{2} \Lm_{i \gamma} \ftC_\alpha{}^\gamma.
\end{align}
To go from the first to second line here, we used the fact that $\langle c_\alpha c^\gamma\rangle \propto \int d \Omega e^{- 2 \pi \bs \cdot (\bb_\alpha \freq + \bb_\beta \freq')/c} |A(\bs)|^2$ which, as discussed above, integrates to zero for $b_{\alpha/\beta} \gtrsim \dAnt$. 

\subsection{Derivation of Equation~\ref{eq:Simplification4}}
Following the same steps used for equation~\ref{eq:Simplification3}, 
\begin{align}
\langle \fc_\alpha \fphi_i^* \rangle & = \frac{S_0^{-1}}{2i}\int d \freq d \freq' e^{-2 \pi i (\freq - \freq')\tau}\Pm_{i \gamma}\langle c_\alpha (c^{\gamma} - c^{\gamma*})\rangle \nonumber \\
& \approx \frac{iS_0^{-1}}{2} \Pm_{i \gamma} \int d \freq d \freq' e^{-2 \pi i(\freq-\freq')\tau} \covC_\alpha{}^\gamma(\freq,\freq') \nonumber \\
& = S_0^{-1}\frac{i}{2} \Pm_{i \gamma} \ftC_\alpha{}^\gamma.
\end{align}

\subsection{Derivation of Equation~\ref{eq:Simplification5}}
We may show this last identity by expanding the real and imaginary components of $c$.
\begin{align}
\langle \feta_i' \fphi_j^{\prime*} \rangle & =S_0^{-2} \int d \freq d \freq' e^{2 \pi i (\freq-\freq')\tau} \Lm_{i \gamma}\Pmt_{\delta \j}\langle \re(c^{\gamma}) \im(c^{\delta})^* \rangle \nonumber \\
& =\frac{S_0^{-2}}{4i} \int d \freq d \freq' e^{2 \pi i (\freq-\freq')\tau} \Lm_{i \gamma}\Pmt_{\delta \j}\langle (c^\gamma + c^{\gamma*})(c^\delta - c^{\delta*}) \rangle \nonumber \\
&\approx 0
\end{align}
We obtain the last line approximately equal to zero due to the fact that $\langle (c^\gamma + c^{\gamma*})(c^\delta - c^{\delta *})\rangle = \langle c^\gamma c^\delta \rangle + \langle c^{\gamma*} c^{\delta*} \rangle + \langle c^{\gamma} c^{\delta*} \rangle - \langle c^{\gamma*}c^{\delta}\rangle$. The first two terms evaluate to zero since they involve integrals over $e^{\pm 2 \pi i \bs \cdot (\bb_\gamma\freq + \bb_\delta \freq')}$ and the last two terms are equal to each other so they subtract to give $0$. 

\section{Components of $\Lm$ and $\Pm$ for Non-Redundant, Uniformly Weighted Calibration Solutions}\label{app:LambdaSums}
In this section, we derive equations~\ref{eq:LMatrix} and \ref{eq:PMatrix} which are valid when the weights matrix is equal to unity. While of limited applicability, they provide us with insight into the scaling of modeling noise with properties of the source catalog and array and allow us to identify the degree to which any visibility covariance contributes to the covariances of gain solutions.

\subsection{Equation~\ref{eq:LMatrix}}\label{ssec:LMatrix}
We wish to evaluate 
\begin{equation}
\Lm_{i \gamma} = [( \bA \bAt)^{-1} \bAt]_{i \gamma}.
\end{equation}
We start with $(\bA \bAt)_{i j}$. Evaluating this matrix product for a non-redundant array is straightforward since each element is given by the dot-product of the $i^{th}$ column of $\bA$ with the $j^{th}$ column. Since a given column is equal to unity at the indices of visibilities in which that antenna participates in and zero otherwise, the dot product of columns is equal to $\nAnt-1$ if $i=j$ and equal to unity if $i\ne j$.
\begin{equation}
\bA \bAt =  \begin{pmatrix} \nAnt-1 & 1 & 1 & \hdots & 1 \\
												1 & \nAnt-1 & 1 & \hdots & 1 \\
												1 & 1 & \nAnt-1 & \hdots & 1 \\
												\vdots & \vdots & \vdots & \ddots & \vdots \\
												1 & 1 & 1 & \hdots & \nAnt-1
												 \end{pmatrix}.
\end{equation}
$\bA \bAt$ can be decomposed into the sum of a diagonal matrix and a matrix formed from an  outer product,
\begin{align}  
\bA \bAt &= \begin{pmatrix} \nAnt-2 & 0 & 0 & \hdots & 0 \\
												0 & \nAnt-2 & 0 & \hdots & 0 \\
												0 & 0 & \nAnt-2 & \hdots & 0 \\
												\vdots & \vdots & \vdots & \ddots & \vdots \\
												0 & 0 & 0 & \hdots & \nAnt-2
												 \end{pmatrix} \nonumber \\
				&+ \begin{pmatrix} 1 & 1 & 1 & \hdots & 1 \\
												1 & 1 & 1 & \hdots & 1 \\
												1 & 1 & 1 & \hdots & 1 \\
												\vdots & \vdots & \vdots & \ddots & \vdots \\
												1 & 1 & 1 & \hdots & 1
												 \end{pmatrix},
\end{align}
and thus can be inverted using the Sherman-Morrison formula.
\begin{equation}
(\bA \bAt)^{-1}_{ij} = \begin{cases} 
									\frac{1}{\nAnt-2} \left[ 1- \frac{1}{2(\nAnt-1)} \right] & i = j \\
									\frac{-1}{2(\nAnt-1)(\nAnt-2)}  & i \ne j.
									\end{cases}
\end{equation}
We can now evaluate $(\bA \bAt)^{-1}_{ij} (\bAt)^j{}_\gamma$ which is the sum of the entries in the $i^{th}$ row of $(\bA \bAt)^{-1}$ that correspond to antennas that participate in the $\gamma^{th}$ baseline. If $i \in \gamma$, we add an entry in $(\bA \bAt)^{-1}$ where $i=j$ to an entry where $i \ne j$. If $i \not \in \gamma$ we add two entries in $(\bA \bAt)^{-1}$ where $i \ne j$. For these two cases we get,
\begin{equation}
\Lm_{i\gamma} = [(\bA \bAt)^{-1} \bAt]_{i \gamma} = \begin{cases}
																							\frac{1}{\nAnt-1} &  i \in \gamma \\
																							\frac{-1}{(\nAnt-1)(\nAnt-2)} & i \not \in \gamma,  \end{cases}
\end{equation}
which completes the proof. 
\subsection{Equation~\ref{eq:PMatrix}}\label{ssec:PMatrix}
We begin evaluating $\Pm_{i \gamma} = (\bB \bBt)^{-1} \bBt$ with the product $(\bB \bBt)_{ij}$ which is the dot product of the $i^{th}$ column of $\bB$ with the $j^{th}$ column. Each $i^{th}$ column contains $\nAnt-1$ non-zero rows that are $1$ when the $i^{th}$ antenna is the non-conjugated participant in the baseline and $-1$ when the antenna is the conjugated participant. The last row of $\bB$ is composed entirely of ones. Thus the dot product of any column with itself is $\nAnt$ and the dot product of a column with any other is equal to zero. Thus,
\begin{equation}
(\bB \bBt)_{ij} = \begin{cases} \nAnt &  i = j \\
														0 & i \ne j
														\end{cases},
\end{equation}
whose inverse is trivial. $(\bB \bBt)^{-1}_{i j} (\bBt)^j{}_\gamma$ is the sum of each element of the $i^{th}$ row of $(\bB \bBt)^{-1}$ that participates the $\gamma^{th}$ visibility.  Since $(\bB \bBt)^{-1}$ is diagonal, this sum is only non-zero when $i=j$. If $i$ is the non-conjugated antenna in the visibility, than $\bBt$ multiplies by $1$ and if $i$ is the conjugated antenna in $\gamma$, $\bBt$ multiplies by $-1$. We obtain 
\begin{equation}
\Pm_{i \gamma} = \begin{cases} \frac{1}{\nAnt} & \gamma = (i,\cdot) \\
													-\frac{1}{\nAnt} & \gamma = (\cdot, i) \\
													0							& i \not \in \gamma
													\end{cases},
\end{equation}
completing our proof.

\section{A Simplified Expression for Minimal Accessible Line-of-Sight Modes.}\label{app:KParMin}
In this section, we derive equation~\ref{eq:kParMin} from equation~\ref{eq:CovVisSuperSimple} with the additional assumptions that the array has a Guassian beam with standard deviation $\sigmaBeam=\epsilon \lambda_0 /\dAnt$ and that its antennas are arranged in a Gaussian configuration with standard deviation $\sigmaAnt$. We also ignore the contribution from thermal noise, which integrates down with time, and assume that $\bC = \bR$.  We first compute $\ftC_{\delta\delta}(\tau,\tau)$ for a Gaussian beam
\begin{align}
\ftC_{\delta\delta}(\tau,\tau) &= \sigma_r^2\int d \freq d \freq' \int d \Omega  e^{2 \pi i \tau(\freq-\freq')}e^{-2 \pi i \bb_\delta \cdot \bs (\freq-\freq')/c}|A(\bs)|^2 \nonumber \\
&\approx \sigma_r^2\int d \freq d \freq' \int d \Omega  e^{2 \pi i \tau(\freq-\freq')}e^{-2 \pi i \bb_\delta \cdot \bs (\freq-\freq')/c}e^{-s^2/2\sigmaBeam^2} \nonumber \\
& \approx  \sigma_r^2 B \int d \Delta \freq \int d \Omega e^{2 \pi i \tau \Delta \freq}e^{-2 \pi i \bb_\delta \cdot \bs \Delta \freq/c} e^{-s^2/\sigmaBeam^2} \nonumber \\
&\approx \sigma_r^2 B\frac{c}{b_\delta} \sqrt{\pi} \sigmaBeam \exp \left(-\frac{c^2\tau^2}{b_\delta^2 \sigmaBeam^2} \right).
\end{align}
To derive the last line, we used the flat-sky approximation, letting the angular integral run over infinity. We also approximate the bandwidth as infinite. Thus, 
\begin{equation}
 \langle \ftC_{\delta \delta} \rangle_{i \in \delta} \approx \sigma_r^2 cB \sqrt{\pi} \sigmaBeam  \left \langle \frac{1}{b_\delta} \exp \left(-\frac{c^2\tau^2}{b_\delta^2 \sigmaBeam^2} \right) \right \rangle_{i \in \delta}.
\end{equation}

Since the chromaticity increases monotonically with increasing baseline length and the antennas with the largest numbers of short baselines are at the center of the array, the minimal $\kparmin$ accessible by an interferometer will occur on a short baseline formed from two antennas near the core of the array. With the core antenna positions equal to $r_i \approx 0$ so that $b_\delta =| {\bf r}_i - {\bf r}_k| \approx r_k$, the average of a function of the length of baselines that a core antenna participates in is equal to the average of that function over antenna positions
\begin{equation}
 \langle \ftC_{\delta \delta} \rangle_{i \in \delta} \approx \sigma_r^2 cB \sqrt{\pi} \sigmaBeam  \left \langle \frac{1}{r_k} \exp \left(-\frac{c^2\tau^2}{r_k^2 \sigmaBeam^2} \right) \right \rangle_{r_k}.
\end{equation}
We can compute this average analytically if the antennas are distributed as a Gaussian with standard deviation $\sigmaAnt$. 
\begin{align}
\left \langle \frac{1}{r_k} \exp \left(-\frac{c^2 \tau^2}{r_k^2 \sigmaBeam^2} \right) \right \rangle_{r_k} & = \frac{1}{2 \pi \sigmaAnt^2}\int d^2r_k r_k^{-1} e^{- \frac{c^2 \tau^2}{r_k^2 \sigmaBeam^2} - \frac{r_k^2}{2 \sigmaAnt^2}} \nonumber \\
& = \sqrt{\frac{\pi}{2}} \frac{1}{\sigmaAnt} \exp\left(-\frac{\sqrt{2}c \tau}{\sigmaBeam \sigmaAnt} \right). 
\end{align}
It follows, that for $i$ and $j$ antennas close to the core, the averages over baselines evaluate to
\begin{equation}
\langle \ftC_{\delta \delta} \rangle_{i \in \delta} \approx \langle \ftC_{\delta \delta} \rangle_{j \in \delta} \approx \sigma_r^2 cB \frac{\pi}{\sqrt{2}} \frac{\sigmaBeam}{\sigmaAnt} \exp\left(-\frac{\sqrt{2}c \tau}{\sigmaBeam \sigmaAnt} \right).
\end{equation}
Thus, for two core antennas with Gaussian beams in a Gaussian antenna distribution, the contamination from calibration errors in equation~\ref{eq:CovVisSuperSimple} reduces to
\begin{equation}\label{eq:GaussianVisibilityNoise}
\mathsf{P}_{\alpha \alpha} - \ftC_{\alpha \alpha} \approx  2\sigma_r^2 B\frac{c}{\nAnt} \frac{\pi}{\sqrt{2}} \frac{\sigmaBeam}{\sigmaAnt} \exp\left(-\frac{\sqrt{2}c \tau}{\sigmaBeam \sigmaAnt} \right) 
\end{equation}
The minimal delay where the signal can be measured, $\tau_{\text{min}}$, is set by where the calibration noise passes below the signal. Thus, we obtain $\tau_{\text{min}}$ by setting equation~\ref{eq:GaussianVisibilityNoise}, multiplied by the prefactors in equation~\ref{eq:DelayPS} that convert from Jy$^2$Hz$^2$ to mK$^2h^{-3}$Mpc$^3$ equal to the 21\,cm power spectrum,
\begin{align}\label{eq:MinCondition}
P_{21} &\approx \left(\frac{\lambda_0^2}{2 k_B} \right)^2 \frac{X^2Y}{B_{pp} \Omega_{pp}} \left( \mathsf{P}_{\alpha \alpha}(\tau_{\text{min}},\tau_{\text{min}}) - \ftC_{\alpha \alpha}(\tau_{\text{min}},\tau_{\text{min}}) \right) \nonumber \\
& \approx  \left(\frac{\lambda_0^2}{2 k_B} \right)^2 \frac{X^2Y}{B_{pp} \Omega_{pp}}\sqrt{2}\pi\sigma_r^2 B  \frac{c}{\nAnt} \frac{\sigmaBeam}{\sigmaAnt} \exp\left(-\frac{\sqrt{2}c \tau_{\text{min}}}{\sigmaBeam \sigmaAnt} \right),
\end{align}
and invert it.
\begin{align}
\tau_{\text{min}} & \approx \frac{\sigmaBeam \sigmaAnt}{\sqrt{2}c} \log \left( \frac{\lambda_0^4 X^2 Y \sqrt{2}\pi \sigma_r^2 c \sigmaBeam}{4 k_B^2 P_{21} \Omega_{pp} B_{pp} \nAnt \sigmaAnt }\right) \nonumber \\
& = \frac{\epsilon}{\sqrt{2} \freq_0} \frac{\sigmaAnt}{\dAnt} \log \left( \frac{\lambda_0^4 X^2 Y \sigma_r^2 B \freq_0}{B_{pp} 2 \sqrt{2} k_B^2 \nAnt \epsilon P_{21}}\frac{\dAnt}{\sigmaAnt} \right)
\end{align}
Using the fact that $\kparmin = 2 \pi \tau_{\text{min}} / Y$, we arrive at equation~\ref{eq:kParMin}. 
\begin{equation}\label{eq:kParMinDerived}
\kparmin = \frac{\epsilon\sqrt{2}\pi}{Y\freq_0} \frac{\sigmaAnt}{\dAnt} \log \left( \frac{\lambda_0^4 X^2 Y \sigma_r^2 B \freq_0}{B_{pp} 2 \sqrt{2} k_B^2 \nAnt \epsilon P_{21}}\frac{\dAnt}{\sigmaAnt} \right).
\end{equation}
Checking this approximate expression against direct calculation for arrays with Gaussian beams using equation~\ref{eq:CovVisSimplified} yields an accuracy of $\approx 10\%$, even in arrays that are not strictly Gaussian such as the MWA, LOFAR, and HERA where $\sigmaAnt$ is the standard deviation of the non-Gaussian antenna distribution. 

The primary shortcoming of equation~\ref{eq:kParMinDerived} is that it assumes a Gaussian primary beam which only accounts for the delay at which the contamination from the main-lobe falls beneath he signal.  Since side-lobes can easily enter at the $ \gtrsim5\%$ level, it is possible for them to contaminate the EoR window at much larger $\kpar$ than the $\kparmin$ predicted in equation~\ref{eq:kParMinDerived}. While the contribution of side-lobes for different baselines will fall at different delays and will not add coherently when averaging over the antenna distribution, we can assume that they add directly to obtain a conservative upper bound on when their contribution will affect $\kparmin$.  If the side-lobes added directly in the antenna average, than their contribution to the amplitude of foreground-residuals would be on the order of $f_{sl}^2$ the level of the foreground residuals at zero delay, where $f_{sl}$ is the ratio between the gain of the side-lobe and the gain at bore-sight. A conservative estimate of when side-lobes are at the level of the 21\,cm signal can be obtained by setting the right hand side of equation~\ref{eq:MinCondition} at zero-delay multiplied $f_{sl}^2$ equal to the 21\,cm power spectrum.
\begin{align}
P_{21} &\approx f_{sl}^2 \left(\frac{\lambda_0^2}{2 k_B} \right)^2 \frac{X^2Y}{B_{pp} \Omega_{pp}}\sqrt{2}\pi\sigma_r^2 B  \frac{c}{\nAnt} \frac{\sigmaBeam}{\sigmaAnt} \nonumber \\
& \approx f_{sl}^2 \left(\frac{\lambda_0^2}{2 k_B} \right)^2 \frac{X^2Y}{B_{pp} \pi}\sqrt{2}\pi\sigma_r^2 B  \frac{c}{\nAnt} \frac{\dAnt}{\epsilon \sigmaAnt \lambda_0}
\end{align}
We use this condition to denote the white-dashed region of parameter space in Fig.~\ref{fig:kParMinConfusion} where side-lobes may render the predictions of equation~\ref{eq:kParMin} inaccurate.

\end{document}